**A Machine Learning Closure for Polymer Integral Equation Theory**

*Zhihao Feng[1], Christian T. Randolph[2], Tyler B. Martin[3], Thomas E. Gartner III[2]**

[1] School of Chemical & Biomolecular Engineering, Georgia Institute of Technology, Atlanta, GA 30332, USA

[2] Department of Chemical & Biomolecular Engineering, Lehigh University, Bethlehem, PA 18015, USA

[3] Materials Science & Engineering Division, National Institute of Standards and Technology, Gaithersburg, MD 20899, USA

*Corresponding author contact information:
Thomas E. Gartner III, teg323@lehigh.edu

Abstract:

Polymer reference interaction site model (PRISM) theory, a descendant of Ornstein-Zernike liquid state theory, is a powerful tool to predict the structure and thermodynamics of equilibrium polymer systems, but its accuracy and applicability can be limited in some important cases. Typically, these shortcomings are traced to the analytical closure relationships used to solve the integral equations. Here, we propose a machine learning (ML)-based closure relation trained on a dataset of coarse-grained molecular dynamics simulations of homopolymer melts and solutions. PRISM theory with the ML closure outperforms traditional atomic closures (*e.g.*, Percus-Yevick) in predicting the structure of typical coarse-grained model systems. We also use the ML closure to accurately model the results of small-angle neutron scattering experiments. This ML-enhanced PRISM theory can therefore enable rapid soft materials discovery and design efforts.

Introduction:

Polymers are an extremely challenging class of materials from a modeling and computation standpoint, due to their coupled static and dynamic behavior crossing orders of magnitude in length and time scales[1]. Molecular simulations provide detailed molecular-level information on polymer properties and behavior but can easily reach prohibitive computational cost[1]. Theoretical approaches can provide more rapid predictions and key physical insight[2-5]—often at the cost of microscopic detail—though quantitative theory is often limited in scope or simply nonexistent for many classes of polymeric materials. Data-driven methods provide a potentially powerful alternative for rapid materials design and discovery, though issues with data sparseness, data representation, and processing-dependence of polymer properties make data-driven polymer design a challenging (though rapidly advancing) field[6-10]. As such, computational polymeric materials design and discovery is not fully solvable via any single method, motivating efforts towards combining the best aspects of multiple approaches to achieve this goal.

Polymer Reference Interaction Site Model (PRISM) theory is an integral equation approach to predicting the structure and thermodynamics of isotropic disordered polymer systems that has many potential advantages for rapid materials screening applications[2, 11, 12]. A descendant of the Reference Interaction Site Model (RISM) theory[13, 14], which itself is based on the classical Ornstein−Zernike (OZ) liquid state theory[15, 16], the PRISM governing equation is defined as follows within a liquid-like system.

$$\hat{H}(k) = \hat{\Omega}(k) \cdot \hat{C}(k) \cdot [\hat{\Omega}(k) + \hat{H}(k)] \qquad \text{Eq [1]}$$

Each term is a two-body correlation function matrix, here represented in Fourier space as a function of wavenumber, $k = \frac{2\pi}{r}$, with $r$ as the distance in real space. Each matrix is symmetric, with dimensions set by the number of site types in the fluid (*e.g.*, atoms or monomers along the polymer chain, nanoparticles, solvent molecules, *etc.*); the correlation function between site types $i$ and $j$ is given by the $(i, j)$ component of the matrix. $\hat{H}(k)$ is the intermolecular total correlation function, which describes the spatial correlation between sites within different molecules. $\hat{\Omega}(k)$ is the intramolecular correlation function, which describes correlations between sites within the same molecule; essentially $\hat{\Omega}(k)$ characterizes the "shape" of the molecule. $\hat{C}(k)$ is the direct correlation function, which is formally defined via **Eq [1]** but can be analogized as an effective pair potential between sites. For a homopolymer system (only one site type), these terms reduce to, $\hat{H}(k) =$ $\rho^2 h(k)$; $\hat{\Omega}(k) = \rho\omega(k)$; $\hat{C}(k) = c(k)$ with $\rho$ as the site number density. Typically, $\omega(k)$ is obtained from analytical expressions representing ideal chain models (*e.g.*, Gaussian chains[17]) or single-chain molecular simulations, leaving $h(k)$ and $c(k)$ as unknowns to be solved iteratively by coupling **Eq [1]** with a so-called "closure" function. While the PRISM equation is exact, the closures are approximations, typically derived as expansions of the free energy. Some classic atomic analytical closure functions are Percus−Yevick (PY)[16, 18] and hypernetted chain (HNC)[16, 19], which are defined as follows.

$$c^{PY}(r) = \left[1 - e^{\beta u(r)}\right] \cdot [h(r) + 1] \quad \text{Eq [2]}$$

$$c^{HNC}(r) = h(r) - \ln[h(r) + 1] - \beta u(r) \quad \text{Eq [3]}$$

$\beta = \frac{1}{k_B T}$, $k_B$ is the Boltzmann constant, and $T$ is the temperature. $u(r)$ is the pairwise interaction potential between sites in the fluid.

The solved PRISM equation enables predictions of useful structural and thermodynamic properties (*e.g.*, pair correlation functions $g(r) = h(r) + 1$, structure factor $s(k) = \rho^2 h(k) + \rho\omega(k)$, Flory-Huggins $\chi$, and more) for disordered isotropic macromolecular liquids, such as homopolymer melts[20, 21] and solutions[22], polymer blends[22, 23], copolymers[24, 25], polymer nanocomposites[26-33], and many others[34-43]. A key advantage of PRISM theory is its computational efficiency—typical solution times are on the order of seconds-to-minutes on standard desktop hardware, even for experimentally-relevant molecular weights, compared with days to weeks on a supercomputer for a particle-based simulation. However, PRISM theory can face applicability challenges under certain conditions. For instance, the accuracy of PRISM predictions with traditional atomic closures typically decreases as the strength of attractive interactions increases, and numerical solution of PRISM theory becomes more difficult to converge near phase boundaries. Typically, the analytical closure functions are considered the key limitation of PRISM theory and the major source of these challenges. Further, the correct closure for a given application is nontrivial and often chosen in an *ad hoc* fashion, as no single closure performs best in all cases[12].

A major step in closure development for PRISM theory was the so-called 'molecular' closures, which include the $\omega(k)$ in the closure expression and fixed known inaccuracies with atomic closures[44, 45]. However, molecular closures present implementation challenges due to the complex convolution integrals needed in the numerical solution pathway, and to the authors' knowledge, there is no publicly available PRISM theory codebase that features a working molecular closure implementation. The current state-of-the-art in PRISM closure development is the two-molecule theory, which reduces a many-chain problem down to a two-chain system in a self-consistently determined solvation field[43]. This approach resulted in a greatly improved prediction of polymer structure relative to reference many-chain simulations compared to the standard PY closure. However, the nested solution loop requires Monte Carlo simulations at every iteration step, and residual deviations from the simulation reference remain, attributed primarily to the approximate HNC form of the solvation potential[43]. As such, the search for accurate and stable analytical closures is ongoing in liquid state theory[46], though no clear best closure has yet been identified[47].

Some progress has been made to augment liquid state theory via data-driven approaches. For instance, a data-driven closure[48] exceeded the performance of some classic atomic closures (*e.g.*, PY, HNC, and the original Verlet closure[49]) when used within an inverse design framework to develop intermolecular potentials from structural data. However, this closure was not used with a full integral equation approach to directly predict liquid structure. Another example of data-driven augmentation of liquid state theory used evolutionary optimization to tune the values of a three-parameter Modified Verlet (MV) closure[50]. However, the optimized MV closure only performed better than the standard atomic closures in some limited cases. There have also been attempts to use physics-informed ML algorithms to solve the OZ equation subject to a given analytical closure[51, 52]. Another option is to bypass the integral equation framework entirely and train regression models to directly predict $g(r)$ based on the state point in Lennard-Jones (LJ) and hard-sphere fluids[53]. Despite its effectiveness in a low-dimensional factorial space (*e.g.*, the volume fraction is the only relevant feature for hard-sphere fluids), this method might be challenging to extend to more complex systems impacted by multiple state variables. Recently, a deep operator network was used to predict the bridge function, which was used as a correction term to the HNC closure within an integral equation calculation[54]. This approach provided excellent predictions for the structure of the LJ, Mie, and hard-sphere fluids. However, in all of these cases only single-site fluids were explored, and macromolecules were not considered.

In this work, we developed an ML-based closure for PRISM theory using feed-forward neural networks to address the key shortcomings with analytical closures. The “ML closure” provides a mapping which enables prediction of the direct correlation function $[c(k)]$ as a function of the intermolecular total correlation function $[h(k)]$ and characteristics of the polymer system and thermodynamic state. Our approach can thus function as a drop-in replacement for the traditional analytical closures in the PRISM solution loop, despite the lack of an interpretable analytical form. We also developed an ML model to predict the intramolecular correlation function $[\omega(k)]$ for any state point. We compared the predictions of our approach to simulation results for two-body structural correlation functions, isothermal compressibility ($\kappa_T$) and the average nearest-neighbor contact distance. We also applied the ML closure to model experimental data obtained via small-angle neutron scattering (SANS) experiments. Predicting and understanding the structure and thermodynamics of disordered fluid systems has been a vital problem of interest to theoretical physicists for more than a century, and our ML-based closure performs significantly better than traditional atomic closures in most of our test cases. Our approach connects theory, simulations, and ML to develop a physical model that combines the advantages of each technique into a holistic framework.

Methods:

Molecular Dynamics Simulations:

To generate data for our model training process, we used the LAMMPS[55] molecular simulation software[56] to perform coarse-grained (CG) molecular dynamics (MD) simulations of a standard bead-spring homopolymer in an implicit solvent. We performed isothermal-isochoric MD simulations with varying chain lengths, polymer-polymer interactions, and polymer concentrations spanning from semidilute solutions to polymer melts. We performed simulations at all combinations of the $(N, \varepsilon, \rho)$ values listed in **Table 1**, where $N$ is the chain length of each polymer; $\varepsilon$ is the well-depth of the LJ potential[57]; $WCA$ is the Weeks-Chandler-Andersen potential[58]; $\rho$ is the number density of monomer segments. The range of monomer-monomer attraction strengths spans good solvent conditions (purely repulsive interactions) to moderately poor solvents (monomer-monomer attraction of $0.5k_BT$).

**Table 1. Model Training State Points from MD Simulations**

| $N$ | $\varepsilon$ | $\rho$ |
|---|---|---|
| [20, 50, 100] | [WCA, 0.05, 0.1, …, 0.45, 0.5] | [0.2, 0.25, …, 0.75, 0.8] |

(1) Initial configurations. CG bead-spring chains with monomer diameter $d = \sigma = 1$ (in LJ reduced units) were randomly placed in a cubic simulation box with periodic boundary conditions. For chain length, $N$ (e.g., 20, 50, and 100), the systems contain 6400, 12500, and 21600 beads, respectively, at the number density $\rho$ listed in **Table 1**.

(2) Simulation setup. All nonbonded interactions were described using the LJ potential,

$$U(r) = 4\epsilon\left[\left(\frac{\sigma}{r}\right)^{12} - \left(\frac{\sigma}{r}\right)^{6}\right] \qquad \text{Eq [4]}$$

with a cutoff of $2.5\sigma$ and well-depth, $\varepsilon$, as given in **Table 1**. For the purely repulsive WCA interactions, the LJ potential was truncated at $r = 2^{\frac{1}{6}}\sigma$. All bonded interactions were described using a harmonic potential,

$$U_{bond}(r) = k_{bond}(r - r_0)^2 \qquad \text{Eq [5]}$$

where $k_{bond} = 200\ \varepsilon/\sigma^2$ and $r_0 = 1\sigma$.

(3) Equilibration. Followed by a short energy minimization to relax bead overlaps from the random initial configuration, simulations were performed for $7.0 \times 10^5\tau$ in the isothermal-isochoric ensemble (NVT) using the LAMMPS Nosé-Hoover thermostat[59-61] with a damping parameter, $\tau_{damp} = 1.0\tau$ and a time step size of $0.005\tau$, at a constant temperature of $T^* = 1.0$. At lower densities, up to five replicates were used per state point to improve statistics. The final $4.5 \times 10^5\tau$ were used for data production and analysis.

For each state point, we computed the monomer-monomer intermolecular pair correlation function $[g(r)]$, the intermolecular total correlation function $[h(k)]$, the intramolecular correlation function $[\omega(k)]$, and the direct correlation function $[c(k)]$ (see the Supplementary Material (SM) for details of the MD data analysis). As PRISM theory is only valid in isotropic single-phase systems, some state points that exhibited phase separation (identified via divergence in the low-*k* limit of the monomer-monomer structure factor) were removed from the dataset, resulting in 395 state points for model development.

Basis Function Expansion:

We made use of a basis function expansion of the $h(k)$ and $c(k)$ obtained from the simulation data to reduce the dimensionality of our dataset. We used the solutions of the quantum harmonic oscillator (QHO) wave function as basis functions, which we selected due to the similarity of the overall shape of the QHO basis set to the $h(k)$ and $c(k)$ curves, namely that all functions show their largest magnitude near $k = 0$ and asymptote to zero at high $k$ (**SM Figures S1-S2**). This basis function expansion also ensures that the predicted $c(k)$ is smooth. By fitting both $h(k)$ and $c(k)$, we optimized parameters (*e.g.*, the set of energy levels used and the initial angular frequency guess) included in the QHO basis function expansion. Full details of the QHO fitting procedure are provided in the SM. We used 60 energy levels with an initial angular frequency guess of 1e$^{-3}$ s$^{-1}$ in the QHO expansion, which allowed our basis functions to capture both low-$k$ behavior and the high-$k$ oscillations around zero (**SM Figure S2**). We then fit the angular frequency ($w$) and linear coefficients ($C_q$) to best represent each state point's $h(k)$ and $c(k)$, respectively.

Neural Network Closure:

We then developed a feed-forward neural network (NN) model to map the QHO features of $h(k)$ to the QHO features of $c(k)$. We also included, as input features, some details of the state point, namely $N$, $\varepsilon$, $\rho$, and a flag to distinguish whether the system exhibited attractive (LJ) or purely repulsive (WCA) interactions [$LJ_{flag}$]. We also represent WCA interactions as $\varepsilon = 0$ in the feature set. We initially attempted to optimize the parameters of the NN by minimizing a loss function based on the error in the QHO features for $c(k)$. We found that slight prediction errors in $w$ and $C_q$ led to significant errors in the final $c(k)$ prediction after converting it from the QHO features. Therefore, we chose a modified loss function that optimizes directly on errors in $c(k)$ as follows:

$$Loss = \frac{1}{N_s}\sum_{i}^{N_s}\sum_{j}^{N_k}\left|[k \cdot c(k)]_{true,\,j} - [k \cdot c(k)]_{pred,\,j}\right|_i \qquad \text{Eq [6]}$$

This customized loss function reflects the mean sum of absolute residuals (MSAR) between the predicted $c(k)$ and simulation $c(k)$, scaled by $k$ to ensure that small-magnitude oscillations at high-$k$ are captured. $N_k$ and $N_s$ indicate the length of the correlation function vectors and the total number of state points in the dataset, respectively. The advantages of using $k \cdot c(k)$ instead of $c(k)$ in the loss function are shown in **SM Figure S3**.

To ensure robustness, we developed a five-fold ensemble model. For each sub-model in the ensemble model, we randomly partitioned the dataset into a different 80:20 train:test split. Each sub-model uses a unique feature scaler (ensuring that all features are on a similar scale) due to the random data partition. Then, we averaged the prediction of $c(k)$ from each of the five individually trained sub-models to produce the ensemble solution.

We optimized the neural network (NN) architecture with three hidden layers containing 65 nodes each using *TensorFlow*[62] and a random search strategy[63]. The detailed architecture is summarized in **Figure 1**. We searched across hyperparameters including activation functions, number of nodes in each hidden layer, optimizer type, and training epochs. While a mixed-activation architecture (e.g., *Tanh*, *Softplus*, and *Swish*) achieved slightly better performance than using *ReLU* in all layers, it introduced higher variance among sub-models in the ensemble. To reduce this variance and improve consistency, we adopted *ReLU* as the only activation function. Each sub-model was trained using the *Adam* optimizer[64], a batch size of 1, shuffled input data, and 400 training epochs.

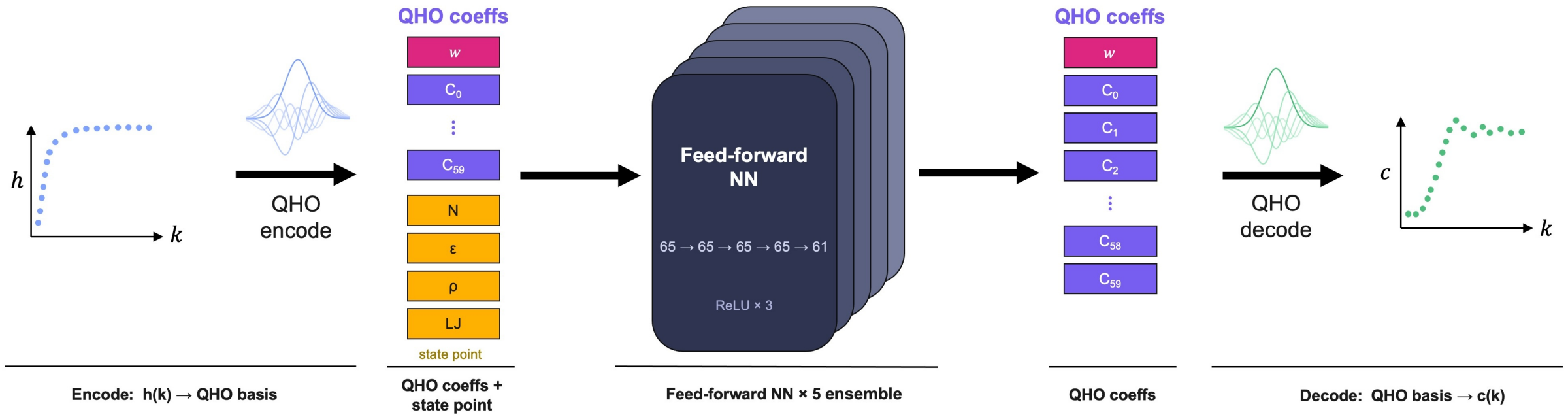

**Figure 1. Architecture of the ML closure.** $h(k)$ is encoded and decomposed to the quantum harmonic oscillator (QHO) basis set, resulting in a vector of an angular frequency ($w$) and 60 coefficients ($C_q$). These QHO features, together with the state point ($N, \varepsilon, \rho, LJ_{flag}$), form the input layer of a feed-forward neural network with three ReLU hidden layers, trained as a five-fold ensemble. The output layer consists of the QHO features of $c(k)$, which are decoded back to $c(k)$ to evaluate the loss function.

Development of $\omega(k)$ Predictor:

For the state points contained in the model training and validation datasets, for which we already performed explicit simulations, we use the simulation-derived $\omega(k)$ for all PRISM calculations (calculated as described in the SM) to more directly isolate the performance of the closure. However, to enable application of the ML closure to state points outside the training dataset, we also developed a model to directly predict $\omega(k)$ from the state point (*e.g.*, $[N,\ \varepsilon,\ \rho,\ LJ_{flag}]$) without additional simulations. For training, each simulated $\omega(k)$ was decomposed using the QHO basis expansion as described above. A feed-forward NN (the architecture is the same as illustrated in **Figure 1**) was then trained on the input features using a single dataset split and rescaling step. The loss function was defined as the MSAR between the predicted and the simulation $\omega(k)$ as follows:

$$Loss = \frac{1}{N_s}\sum_{i}^{N_s}\sum_{j}^{N_k}\left|[\omega(k)-1]_{true,\,j}-[\omega(k)-1]_{pred,\,j}\right|_i \qquad \text{Eq [7]}$$

This predictor was trained using the *Adam* optimizer[64], a batch size of 1, shuffled input data, and 300 training epochs. The effectiveness of this model can be seen in **SM Figures S4 and S5**, in which the difference between the predicted and simulation $\omega(k)$ was negligible.

Numerical Solution Scheme:

After producing the trained NN closure, we integrated it within a Picard iteration loop to solve the PRISM equation for any state point (*i.e.*, set of $(N,\varepsilon,\rho)$) enclosed in the dataset boundaries (**Figure 2**). First, we select an initial guess of $h_0(k) = 1$ for all *k*. Then, we apply the NN closure using the state point and $h_0(k)$ to predict $c_1(k)$. We then apply the PRISM equation (**Eq [1]**) to calculate a new $h_1(k)$ using the predicted $c_1(k)$ and the relevant $\omega(k)$ for that state point. We then take $h_1(k)$ as the new input to the NN closure and iterate until $\frac{1}{N_k}\sum_k[\gamma_i(k)-\gamma_{i-1}(k)]^2$ falls below a specified tolerance. Here, the indirect correlation function, $\gamma(k)$, is defined as $h(k)-c(k)$. We found an initial guess of $h_0(k) = 1$ to be sufficient for stable convergence at the vast majority of state points considered in this work; more physically-informed initial guesses (*e.g.*, solutions from a hard-sphere reference system) may accelerate convergence but were not considered here. To balance accuracy and computational efficiency, the number of Picard iterations was capped at 100, with a convergence tolerance of $10^{-4}$. All standard PRISM theory calculations (without the NN closure) were performed using the Python-based pyPRISM software[12].

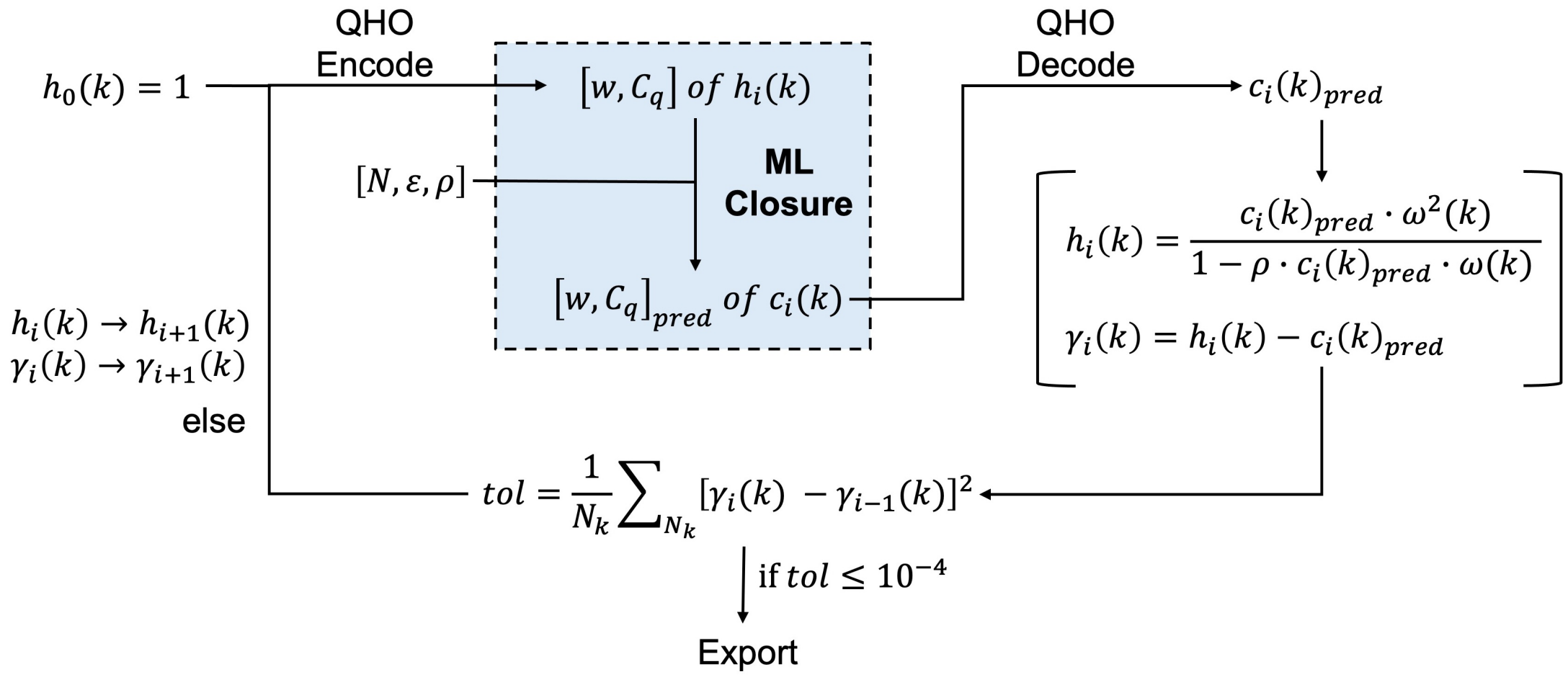

**Figure 2. Flow chart for ML-PRISM numerical solution.** Starting from the initial guess $h_0(k) = 1$, each Picard iteration $i$ encodes $h_i(k)$ into its QHO features $[w, C_q]$, infers the QHO features of $c_i(k)$ at the state point $(N, \varepsilon, \rho)$, and decodes them to $c_i(k)_{pred}$. The PRISM equation then returns an updated $h_i(k)$ using $\omega(k)$ for the state point (from simulation or the $\omega(k)$ predictor), and $\gamma_i(k)$ is then computed. The loop repeats until the tolerance is reached, and the converged solution is exported.

Results:

We note that the NN closure model training process is fully outside the PRISM solution loop, such that PRISM with the ML closure has similar time-to-solution as traditional analytical closures. To examine the performance of our model, we validated its prediction performance at 10 randomly selected state points excluded from the initial training dataset (validation state points).

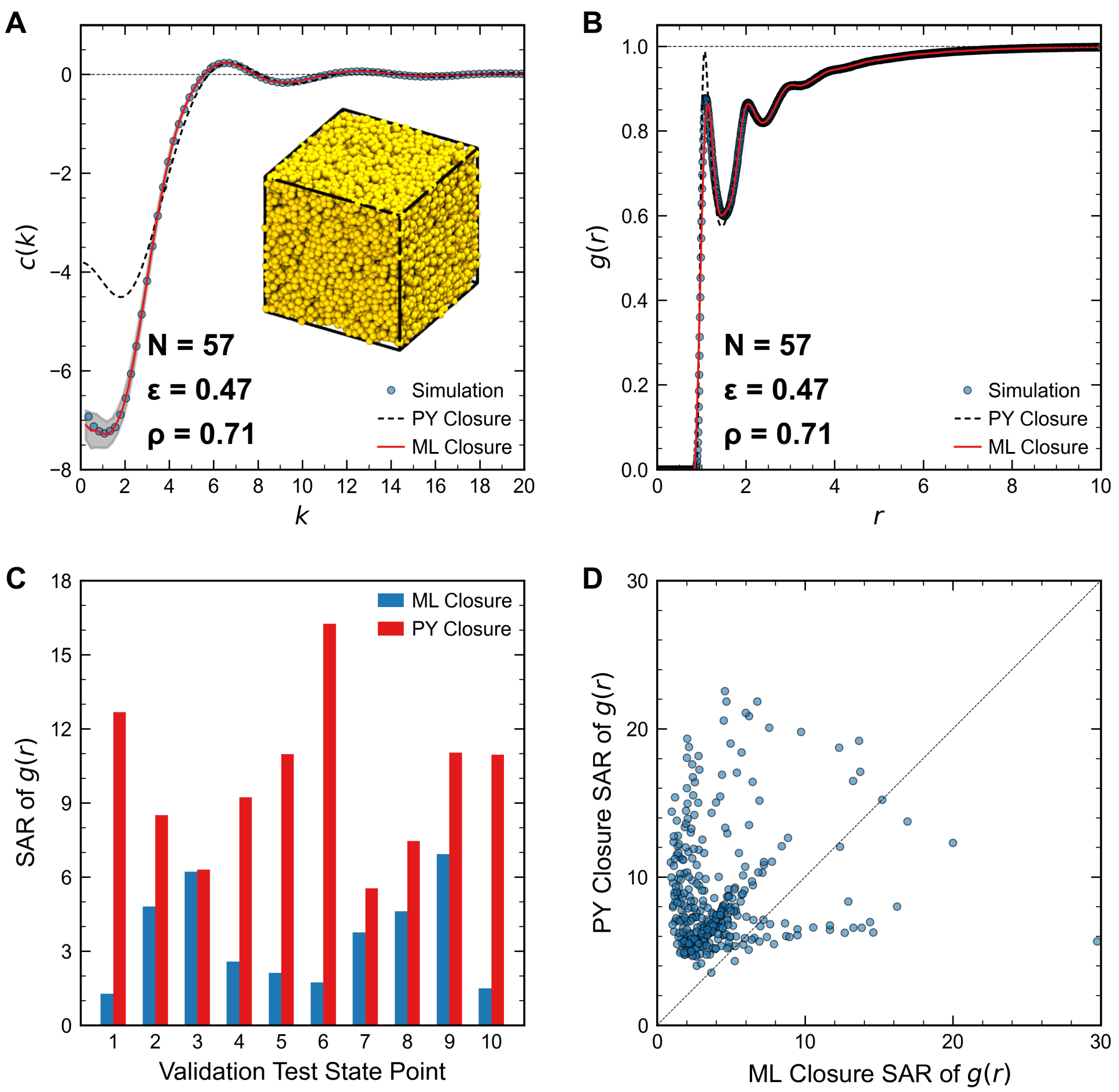

**Figure 3. Model benchmarks.** (A) Direct correlation function, $c(k)$, with the inter-model standard deviation (gray shaded area) and (B) intermolecular pair correlation function, $g(r)$, predicted by the ML closure (red solid curve), the PY closure (black dashed curve), and the simulation reference (blue points) at the state point $N = 57$, $\varepsilon = 0.47$, $\rho = 0.71$ (inset configuration rendered using VMD)[65]. (C) The sum of absolute residuals (SAR) of $g(r)$ at 10 validation state points for the ML closure (blue) and PY closure (red). (D) The SAR of $g(r)$ at all training-set state points for the ML and PY closures.

As shown in **Figures 3A** and **3B** for a randomly chosen state point not in our training dataset, the ensemble solution was obtained by averaging $c(k)$ from the sub-models with a given $N$, $\varepsilon$, $\rho$, $LJ_{flag}$. Note that for the PRISM calculations presented in **Figure 3**, we use the $\omega(k)$ derived from simulation for each state point to isolate the performance of the closure. The consistent agreement among the sub-models can be observed in the uncertainty of the ensemble model, quantified by the standard deviation across sub-model predictions (see SM). Compared to the PY closure, the ML closure ensemble model more closely followed the simulation-derived $c(k)$ across both low- and high-$k$ regimes. The direct correlation function in real space is shown in **SM Figure S6**, which shows the same overall trend as its Fourier-space counterpart and indicates that, for this particular state point, the deviations between the PY closure and the simulation reference are primarily short-ranged. In terms of pairwise structure, the ML closure prediction for $g(r)$ achieved a significantly improved agreement with the simulation. For 10 randomly selected validation state points, the ML closure outperformed the PY closure in terms of the sum of absolute residuals (SAR) of $g(r)$ between the simulation and the PRISM theory prediction, as shown in **Figure 3C**. The ML closure obtained more accurate solutions than PY in 91 % of training state points **(Figure 3D)**. Further, 53 % of training-set state points saw a >50 % improvement in SAR, and 18 % of state points showed a >80 % improvement. We also compared the ML closure to the atomic HNC closure, which performed worse than ML at all validation state points (**SM Figure S7**). We note that the PRISM theory numerical solution loop with the ML closure failed to converge at three training-set state points due to proximity to the phase separation boundary. This compares to 45 failed convergences with the PY closure, possibly indicating that the ML closure is more robust near phase boundaries. We also investigated the impact of dataset size on the ML closure performance. As shown in **SM Figure S8**, approximately 150 training-set state points (40 % of total available simulated state points) are sufficient to train an accurate ML closure for this system.

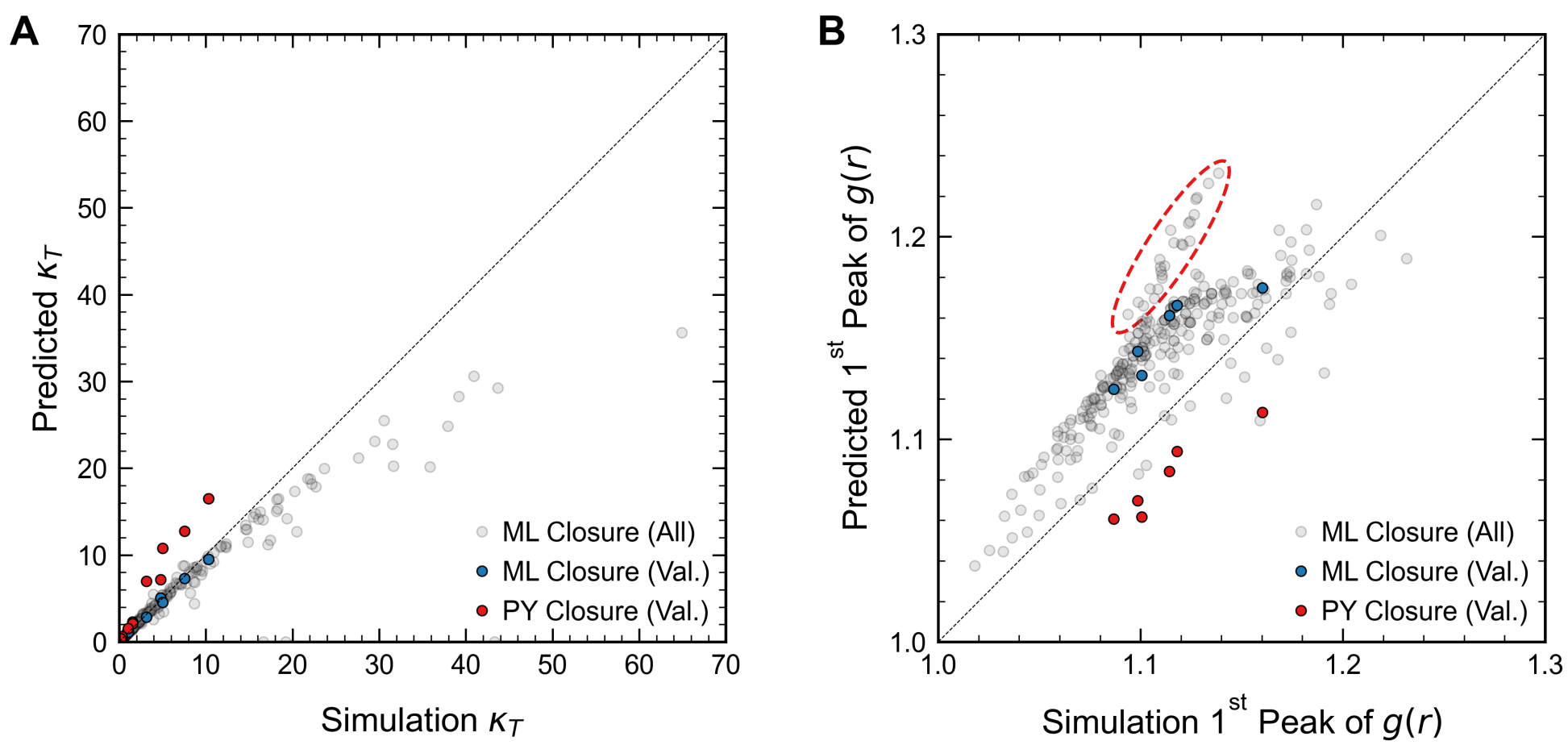

**Figure 4. Isothermal compressibility, $\kappa_T$ (A), and average nearest-neighbor contact distance (B).** Gray symbols represent the ML closure predictions at the training-set state points. Blue and red symbols show the ML and PY closure predictions at the validation set state points, respectively. The red dashed circle in (B) indicates state points that include purely repulsive interactions.

Besides structural correlation functions, we can also explore other structural and thermodynamic quantities. **Figure 4A** shows the isothermal compressibility, $\kappa_T$, calculated for all training and validation state points. For **Figure 4**, we again use only the $\omega(k)$ derived from simulation for each state point. With the converged ensemble solution of $c(k)$, we computed $\kappa_T$ from the structure factor, $s(k)$, with $\kappa_T = \frac{s(k=0)}{\rho k_B T}$ [16]. We approximate $s(k=0)$ with $s(k_{min})$, where $k_{min} = \frac{2\pi}{L}$ is the lowest possible wavenumber sampled in our finite-sized system of side length $L$. While this approach may result in a systematic underprediction of $s(k=0)$ (and thus $\kappa_T$) for the highly compressible state points near the phase boundary, we apply the same procedure for both simulation and theory, likely rendering **Figure 4A** not significantly affected by the specific $s(k \rightarrow 0)$ extrapolation method. Further, we note that there are multiple routes to calculate compressibility, including taking the derivative of free energy or pressure with respect to volume. The resulting $\kappa_T$ may change somewhat as a function of the particular method chosen, particularly for PRISM theory, but we again select the $s(k=0)$ route to maintain consistency between theory and simulation. The ML closure predicted $\kappa_T$ reasonably accurately with respect to the simulation results in most cases; we observed larger deviations near the phase-separation boundary (*i.e.*, state points with larger density fluctuations). We note that the ML closure was more accurate than PY in capturing the magnitude of density fluctuations, except for state points with very low $\kappa_T$, for which the ML and PY predictions were similarly accurate. The average nearest-neighbor contact distance is presented in **Figure 4B**, defined as the $r$ position of the first peak of the intermolecular $g(r)$. For some state points at low $\varepsilon$ and $\rho$ there is no clearly defined first peak—those state points are not included in **Figure 4B**. The ML closure achieved reasonable agreement compared to the simulation reference, although most state points showed a slight overprediction of the nearest-neighbor distance, corresponding to an underprediction of the pressure as a function of density and interaction strength. We note a cluster of state points (circled in red) that indicates the ML closure's lower prediction accuracy for the WCA potential at high and medium densities, and the corresponding strength of the PY closure in this regime. This is not entirely unexpected given that atomic closures often perform best for repulsive hard-sphere systems. Because the PRISM numerical solution loop involves repeated transforms between real- and Fourier-space, we hypothesize that small errors in the high-$k$ oscillation behavior from the QHO basis function may be one source of these systematic errors. In future work, we will explore other possible basis function sets to test this idea. In contrast to the ML closure, the PY closure slightly underestimates the nearest-neighbor contact distance in all cases.

Considering both structural and thermodynamic information, the ML closure outperforms the PY closure in the vast majority of cases explored for this homopolymer test system. However, we note two possible limitations/challenges of our initial ML closure implementation: (1) While the ML closure appears to be more numerically robust than PY near a phase boundary (see above), it may be less accurate under such conditions. Near a phase boundary, the ML closure prediction of $c(k)$ tends to overestimate the strength of correlations relative to the simulation reference. For instance, for the state point $N = 93$, $\varepsilon = 0.33$, and $\rho = 0.3$, which is on the edge of the single-phase region as shown in **SM Figure S9**, the ML closure failed to converge. When we decreased $\varepsilon = 0.33$ to $\varepsilon = 0.23$, the convergence was successful. However, we anticipate that these challenges with the ML prediction may be mitigated by including additional training data near to the phase boundary, rather than the simple grid we used in the present work. (2) The ML closure is less accurate than the PY closure for systems exhibiting purely repulsive (WCA) interactions at medium-to-high densities. We partially attribute this behavior to our representation of the WCA state points at the extreme edge of the $\varepsilon$ axis in our dataset (*i.e.*, $\varepsilon = 0$ for WCA), which may be less accurate compared to interpolation (*i.e.*, $\varepsilon > 0$ for LJ). The sharp difference between the intermolecular $g(r)$ for WCA vs. weak LJ attractions is shown in **SM Figure S10**, which comes about due to the relative softness of the repulsive part of the LJ potential when $\varepsilon$ is small. This observation prompted the addition of the $LJ_{flag}$ to the feature set. Including $LJ_{flag}$ caused a small improvement in the ML closure predictions for the WCA cases (**SM Figure S11**), though the ML closure was still not as accurate as the PY closure under purely repulsive conditions. Possibilities for addressing these limitations are given in the Discussion section below.

To demonstrate possible applications of ML-PRISM outside of simple CG systems, we explored ML-PRISM's ability to act as a model for the interpretation of SANS experiments. We measured the SANS intensity, $I(q)$, of solutions of linear polystyrene (PS) dissolved in p-xylene at polymer volume fractions, $\phi$, of 0.09 and 0.18 (see the SM for full experimental details). Then, we calculated $I(q)$ from our ML-PRISM approach to find the best set of CG model parameters to reproduce the density fluctuations reflected in the experimental $I(q)$. Note that we use $q$ to denote the SANS wavenumber to highlight the difference in units between the experimental $q$ ($\text{Å}^{-1}$) and the PRISM $k$ ($\sigma^{-1}$ in LJ reduced units); see **Eq [8]** below. To translate between our CG model and the experimental system, we took the CG bead diameter to be 11 Å to approximately match the Kuhn segment lengths of the real PS chain and the CG chain (see **SM**); using an approximate monomer diameter of 3 Å results in approximately 3.6 PS monomers per CG monomer. Thus, to model our experimental PS chains of 22,500 g/mol, we used 60 CG monomers, $N$. The experimental $\phi$ values correspond to $\rho$ = 0.17 and 0.34, respectively, with $\sigma = 1.0$ (see details in the SM). To find the best-fit PRISM theory result to match the SANS $I(q)$, we use $\varepsilon$ and $A$ as fitting parameters. Since we do not have reference $\omega(k)$ data for every possible $\varepsilon$, when comparing to the SANS data we estimate $\omega(k)$ for each $\varepsilon$ using our trained $\omega(k)$ predictor model. For select state points, we confirmed that a short MD simulation provides good agreement with the $\omega(k)$ predictor (**SM Figure S12**). We calculate $I(q)$ from the PRISM theory results as follows.

$$I(q) = B + A \cdot \phi \cdot \left(\rho_p - \rho_s\right)^2 \cdot P(k) \cdot \mathrm{s}(k) \qquad \text{Eq [8]}$$

B is the incoherent background; $A$ is a scale factor that accounts for unit conversions and instrument-specific factors; $\rho_p - \rho_s$ is the difference in scattering length densities between the polymer and solvent, which is $(1.99 \cdot 10^{-6}) - (4.62 \cdot 10^{-6})\ \text{Å}^{-2}$ (determined from contrast variation experiments); $P(k)$ is the form factor amplitude for a sphere, expressed by $[\frac{3\sin(u)-u\cdot\cos(u)}{u^3}]^2$, where $u = k \cdot R$ and $R = \sigma/2$[66]. Using **Eq [8]**, we obtained best-fit matches to the experimental $I(q)$ using both the ML closure and the PY closure by optimizing $\varepsilon$ and $A$.

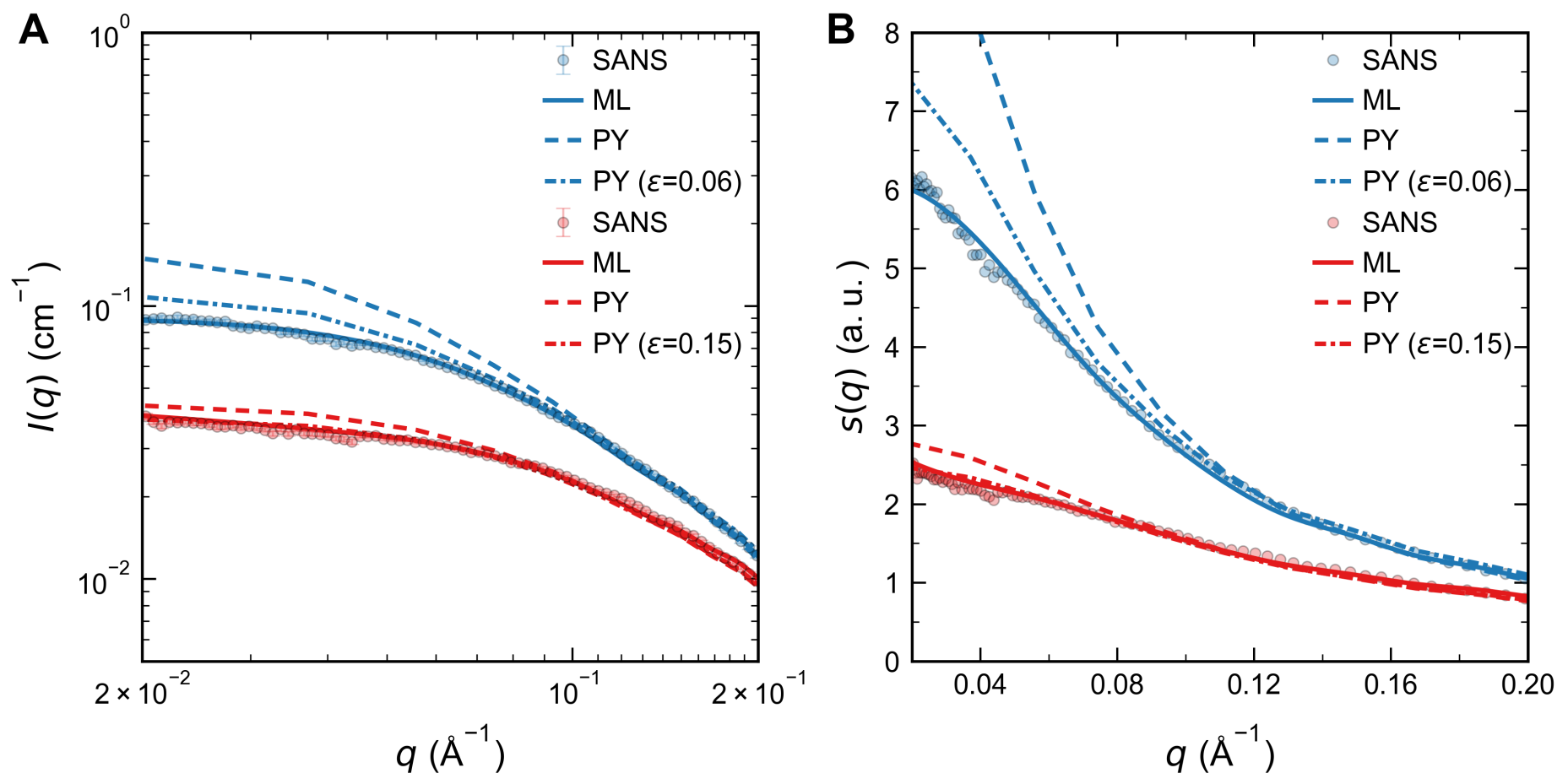

**Figure 5. Modeling SANS Data.** The ML closure (solid curves) and PY closure (dashed curves) for the (A) scattering intensity, $I(q)$, and (B) structure factor, $s(q)$, compared to SANS experiments (points) at $\rho = 0.17$ (blue) and $\rho = 0.34$ (red). The dot-dash curves indicate the best PY closure fits, and the dashed curves indicate the PY closure result using the same $\varepsilon$ as the ML closure. Experimental error bars in (A) are the standard deviation in the measured intensity.

We restricted the fitting procedure to use the same value of $\varepsilon$ for both polymer densities, yielding best-fit inputs of $[\varepsilon, A]$ at $[0.203, 10^{10.375}]$ and $[0.203, 10^{10.100}]$ for $\rho$ of 0.17 and 0.34, respectively. Using this value of $\varepsilon$ within the ML closure provided an excellent match to the SANS data, demonstrating that the PRISM theory results appropriately represent the density fluctuations observed in the experimental system. Further, the ML closure outperformed the PY closure in fitting the SANS data at both densities, as shown in **Figure 5**. Using the same $\varepsilon$ as the ML model (dashed lines in **Figure 5**), PY significantly overpredicted the scattering intensity. To evaluate whether PY could reproduce the density fluctuations under any conditions, we also adjusted $\varepsilon$ to search for the closest match between the PY closure and experiments—this was achieved at $\varepsilon = 0.06$ and 0.15 for $\rho = 0.17$ and 0.34, respectively. While PY appears sufficient to model the SANS intensity for the higher-density system, no value of $\varepsilon$ gave a good correspondence between PY and SANS for $\rho = 0.17$. Thus, the ML closure is a more flexible and accurate choice to model the SANS results, and notably, a single value of $\varepsilon$ was sufficient to model both densities. Importantly, this result provides a possible coarse-graining pathway, allowing one to determine a value of the model's attraction strength that matches the density fluctuations observed by scattering for a given experimental system, which could then be used for future detailed simulations or PRISM theory calculations to provide additional molecular-level insight.

Taken together, these results demonstrate the promising performance of the ML closure in predicting the structure of model systems and interpreting SANS experiments and show significant improvements in performance in both cases compared to the PY closure as a representative atomic closure. We also tested the performance of a more modern closure, namely the Modified Verlet (MV)[29]—for the systems considered here, MV performed similarly to PY and worse than the ML closure (see **SM Figure S13**). We note that all closures used for comparison in this work are atomic closures, which we selected due to their wide application and relative ease of implementation. As such, comparison to molecular closures remains an important benchmark for future studies, particularly for multicomponent systems.

Discussion:

Herein, we used a minimal feature set of $N$, $\varepsilon$, $\rho$, and $LJ_{flag}$, coupled with a simple feed-forward NN architecture to create our ML closure, and we show that this straightforward approach is sufficient to model homopolymer melts and solutions. Given these results, we highlight several important technical findings: (1) simple scalar descriptors of the state point $(N, \varepsilon, \rho)$ encode sufficient information for a neural network model to learn the functional relationship between the total and direct correlation functions. (2) Our model only required ~150 data points and did not significantly improve in accuracy with the inclusion of additional data, which is a strong indicator of generalizability. Such a (relatively) small data demand shows that the approach could potentially be used to create an experimentally-derived closure. (3) The trained ML closure is numerically robust, sustaining Picard iterations for nearly all state points considered (e.g., the ML closure failed in three cases vs. 45 using the PY closure). (4) The ML closure was trained entirely on simulated data, yet it still reproduced experimental SANS scattering intensity of PS/p-xylene and outperformed the PY closure in this task. These traits of the ML closure strongly support that the ML closure is a learned functional rather than a simple regressor.

In further ML closure iterations, we envision expanding to systems that have stronger density fluctuations, multiple components, and/or more realistic polymer models (*e.g.*, polymer blends, copolymers, polymer/particle composites). First, extension to a different model system would be done in much the same way as a traditional PRISM calculation, by adjusting the forms for the intermolecular pair potentials and $\omega(k)$ as needed to reflect the specific polymer model desired. The ML closure would then need to be retrained following the procedure we outline in the present work. Depending on the resulting ML closure performance, one can envision several avenues for development which may provide additional robustness for complex systems. We could encode the physics more directly into the model by using as features the full functional forms of the intramolecular correlation function, $\omega(k)$, and/or the intermolecular potential, $u(r)$, thus creating an ML analog to the molecular closures (e.g., $c(k) \propto ML[\omega(k), u(k), h(k)]$). We may also leverage new data-driven methods designed to handle small datasets[67-69], which may improve performance, especially towards the edges of phase boundaries. We might further improve the applicability of the ML closure by applying transfer learning or multi-fidelity techniques with experimental data. Autonomous experimental platforms, such as the Autonomous Formulation Lab (AFL)[70, 71], could be used to efficiently generate balanced experimental datasets for training. Another avenue could include "physics-informed" adjustments to the loss function[72, 73]. While the current loss function may implicitly include physical information since the simulation $c(k)$ is derived from the PRISM governing equation directly, adding an additional loss term to describe how well the predicted $c(k)$ satisfies the PRISM governing equation may improve the model further. Finally, we note that our ML approach should be thought of as a mapping between correlation functions that performs the same function as a traditional closure but does not yet provide a physically-interpretable analytical expression. Adjustments to the ML approach that may provide such interpretability is a promising avenue for future work. These options may enable robust future extensions to complex polymer systems, but each comes at the expense of additional algorithmic complexity and/or dataset size, potentially decreasing numerical stability or slowing dataset construction and model training tasks.

Conclusion:

In this work, we demonstrated that an ML closure developed with an NN ensemble model and QHO basis function representations of structural correlation functions and integrated into a Picard iteration scheme to solve the PRISM equations can outperform several atomic closures (PY, HNC, MV) in predicting both structural and thermodynamic properties of polymers. Specifically, the ML closure gave more accurate pair correlation functions than tested atomic closures for 10 randomly-selected out-of-dataset state points. Further, the ML closure improved predictions over PY for 91 % of the training-set state points. Although the improvement afforded by ML is not uniform across all of phase space, we identified that the ML closure's gains are largest for the state points with attractive interactions, at low densities, and near to phase-separation boundaries, whereas its accuracy is limited for the WCA potential at medium-to-high densities. Our initial training and test system used coarse-grained homopolymer simulation data as a reference, but we demonstrated that the learned closure can be transferred to interpret data from SANS experiments on polystyrene solutions, where it outperformed the PY closure. For ease of use, we also developed an $\omega(k)$ predictor so that no additional simulations are required to apply the ML closure to a given state point, as demonstrated in the SANS fitting.

This work represents a significant step in combining integral equation theory with data-driven approaches to expand PRISM theory's applicability for polymer materials design, with possible applications in complex formulation screening. Our work also contributes more broadly to the long-standing liquid state theory challenge in the search for accurate general-purpose closure expressions.

Supplementary Material:

See Supplementary Material for full details of MD data analysis, QHO basis function expansion, additional discussion of the ML closure, SANS experimental details, and a self-contained web application to run the ML closure implementation.

Acknowledgments:

Z.F. and T.E.G. acknowledge funding from the School of Chemical & Biomolecular Engineering at the Georgia Institute of Technology. C.T.R. and T.E.G. acknowledge funding from the Department of Chemical & Biomolecular Engineering at Lehigh University. This research was supported in part through research cyberinfrastructure resources and services provided by the Partnership for an Advanced Computing Environment (PACE) at the Georgia Institute of Technology, Atlanta, Georgia, USA.

Author Declarations:

The authors have no conflicts to disclose. T.B.M. and T.E.G. conceived of the project; Z.F. and C.T.R. performed computational research; T.B.M. performed scattering experiments; all authors analyzed data and wrote the paper.

Data and Code Availability:

The ML closure is available via a GitHub repository at https://github.com/fengzhihao77/machine-learning-closure-prism, with installation and usage instructions provided in the repository's README.md file. The LAMMPS trajectory and input files that comprise the training dataset are available upon request.

**Supplementary Material for: A Machine Learning Closure for Polymer Integral Equation Theory**

*Zhihao Feng[1], Christian T. Randolph[2], Tyler B. Martin[3], Thomas E. Gartner III[2]**

[1] School of Chemical & Biomolecular Engineering, Georgia Institute of Technology, Atlanta, GA 30332, USA

[2] Department of Chemical & Biomolecular Engineering, Lehigh University, Bethlehem, PA 18015, USA

[3] Materials Science & Engineering Division, National Institute of Standards and Technology, Gaithersburg, MD 20899, USA

*Corresponding author contact information:
Thomas E. Gartner III, teg323@lehigh.edu

Table of Contents

## Additional Methods Details

### 1. Simulation data analysis

#### 1.1. Intermolecular and intramolecular distance arrays

For each MD trajectory, the pairwise Euclidean distance array was computed for each particle pair, $i$ and $j$ as follows:

$$r_{ij} = \sqrt{(x_i - x_j)^2 + (y_i - y_j)^2 + (z_i - z_j)^2} \quad \text{Eq [S1.1]}$$

The resulting distance array was separated into the intermolecular (off-diagonal) and intramolecular (on-diagonal) subarrays via masking.

#### 1.2. Correlation function calculations

<u>The intermolecular pair correlation function</u>, $g(r)$, was computed from the off-diagonal subarrays using **Eq [S1.2]**. Periodic boundary conditions were implemented in the subarrays using the minimum image convention (e.g., $r_{ij} \leq \frac{L}{2}$, where $L$ is the simulation box length).

$$g(r) = \langle \frac{N_{bin}}{V_{bin} \cdot \rho \cdot N} \rangle \quad \text{Eq [S1.2]}$$

Here, $r$ is the real space range with a maximum value, $r_{max} = \frac{L}{2}$, and a grid spacing, $\Delta r = \frac{L}{2N_k}$, where $N_k$ is the spacing length. $N_{bin}$ is the count of particles in each histogram bin. $V_{bin}$ is the volume of the corresponding spherical shell. The angle brackets indicate averaging over MD production data.

<u>The intramolecular correlation function</u>, $\omega(k)$, was computed from the on-diagonal subarrays using **Eq [S1.3]**[1]. Unwrapped trajectories were implemented in the subarrays.

$$\omega(k) = \langle \frac{1}{N} \sum_{i=1}^{N} \sum_{j=1}^{N} \frac{\sin(kr_{ij})}{kr_{ij}} \rangle \quad \text{Eq [S1.3]}$$

Here, $k$ is the wavenumber in the Fourier space, with a grid spacing, $\Delta k = \frac{\pi}{r_{max}}$.

<u>The total correlation function</u>, $h(k)$, was derived from $g(r)$ using a fast Fourier transform (FFT) in *SciPy* as follows[2]:

$$k \cdot h(k_b) = 4\pi\Delta r \sum_{a=0}^{N_k-1} r \cdot [g(r_a) - 1] \cdot \sin\left[\frac{\pi}{N_k + 1}(a + 1)(b + 1)\right] \qquad \text{Eq [S1.4]}$$

where $a$ and $b$ are integer indices ranging from 0 to $N_k - 1$. $N_k$ was set to be 2048 to optimize the speed of the FFT[3].

The direct correlation function, $c(k)$, was computed from the simulation results via the polymer reference interaction site model (PRISM) governing equation.

$$c(k) = h(k) \cdot \{\omega(k) \cdot [\omega(k) + \rho h(k)]\}^{-1} \qquad \text{Eq [S1.5]}$$

1.3. High-resolution processing for $c(k)$

To improve issues with numerical noise at low-$k$, we adopted a numerical strategy similar to Goodall et al. for calculating Fourier space correlation functions[4]:

(1) Structure factor from PRISM (Method 1).

The structure factor, $s(k)$, was computed from previously calculated $h(k)$ and $\omega(k)$ as follows:

$$s(k) = \rho h(k) + \omega(k) \qquad \text{Eq [S1.6]}$$

It is worth noting that $s(k)$ is normalized by $\rho$. This method suffers from numerical noise at low $k$, but is comparatively computationally efficient.

(2) Structure factor from direct method (Method 2).

Using *Freud*[5], the structure factor was computed directly from MD trajectories by summing vectorial contributions in reciprocal space. These were averaged over MD production data. The direct calculation is less noisy, but computationally expensive.

(3) Data blending.

The structure factor from Method 2 was used up to $k' - 4 \cdot \Delta k$, where $k'$ is the first principal peak location of $s(k)$, and then smoothly merged with the Method 1 data following Goodall et al[4]. To ensure consistency in the values of $k$ across all relevant functions, $\omega(k)$ was recalculated using the *Freud*-defined $k$-space and blended with the $\omega(k)$ obtained by Eq. S1.3 above.

(4) Final computation of $c(k)$.

Using the refined $s(k)$ and $\omega(k)$, $c(k)$ was computed as follows:

$$c(k) = \frac{1}{\rho} \cdot \left(\frac{1}{\omega(k)} - \frac{1}{s(k)}\right) \qquad \text{Eq [S1.7]}$$

This approach yielded high-resolution $c(k)$ with decreased numerical noise while balancing computational efficiency.

## 2. Direct mapping between $h(k)$ and $c(k)$ with a neural network model

As an initial approach, we investigated if a neural network (NN) model, such as a multilayer perceptron (MLP), could directly predict $c(k)$ with given $N$, $\varepsilon$, $\rho$, and $h(k)$. However, unphysical behaviors were observed in the prediction, including large oscillations throughout the entire $k$ range, as shown in **Figure S1**. This result motivated the use of the quantum harmonic oscillator (QHO) basis function expansion to ensure smoothness of the NN prediction.

## 3. Optimization of the quantum harmonic oscillator (QHO) wave function

We employed the energy eigenfunctions solved from the one-dimensional and time-independent Schrödinger equation for the quantum harmonic oscillator (QHO) as basis functions in **Eq [S3.1]**.

$$\psi_n(k) = \frac{1}{\sqrt{2^n n!}} \left(\frac{mw}{\pi\hbar}\right)^{\frac{1}{4}} e^{-\frac{mwk^2}{2\hbar}} H_n\left(\sqrt{\frac{mw}{\hbar}}k\right), n = 0, 1, 2, \ldots, N_q - 1 \qquad \text{Eq [S3.1]}$$

The Hermite polynomials, $H_n(z)$, are defined as:

$$H_n(z) = (-1)^n e^{z^2} \frac{d^n}{dz^n}\left(e^{-z^2}\right) \qquad \text{Eq [S3.2]}$$

Here, $k$ is the input vector. $n$ is the energy level. $m$ is the mass (set to be 1 $[kg]$). $w$ is the angular frequency $[s^{-1}]$. $\hbar$ is the reduced Planck constant. $N_q$ is the total number of energy levels.

We used the QHO eigenfunctions as a basis set to fit the previously calculated curves $h(k)$ and $c(k)$. Specifically, the following procedure was taken:

1. Optimization of $w$.

   For a given $k$ and an initial guess $w_{initial}$, a single eigenfunction $\psi_n(k)$ was generated at an initial guess energy level, $n_{initial}$, and fitted to the data. $w$ was then optimized and applied consistently across all $N_q$ energy levels to construct the full basis. The nonlinear programming objective was solved using *SciPy*[2].

2. Linear combination of energy eigenfunctions.

   Each target curve was represented as a linear combination of $N_q$ energy eigenfunctions,

$$\sum_{0}^{N_q-1} C_q \cdot \psi_n(k) \qquad \text{Eq [S3.3]}$$

3. Grid search.

   A grid search was performed over $N_q$, $n_{initial}$, and $w_{initial}$, to minimize the fitting errors across all training state points for $h(k)$ and $c(k)$, respectively. We found that $N_q = 60$ accurately reconstructed both functions using $w_{initial} = 1e^{-3}$, with $n_{initial} = 3$ for $h(k)$, $n_{initial} = 0$ for $c(k)$, and $n_{initial} = 2$ for $\omega(k)$. Thus, each correlation function was decomposed into one $w$ and 60 coefficients, $C_q$, resulting in 61 features in total.

## 4. Customized loss function

To facilitate learning across all simulation entries, we leveraged the continuous function representation provided by the QHO basis expansion to define a unified Fourier space, $\hat{k}$, with an interval of 0.1 over the range of [0.05, 100]. This unified range offers several advantages:

1. Flexible resolution control. The $\hat{k}$-range can be refined to prioritize the first principal and higher order peaks, since the correlation functions are asymptotic to zero at high $k$-range.
2. Reduced computational cost. Since each simulation's correlation functions are expressed on the same $\hat{k}$-range, the loss function avoids repeatedly calling simulation-specific $k$-range during training.
3. Enhanced sensitivity to local deviations. The sum of absolute residuals (SAR) was employed as the loss metric, which better captured localized deviations, particularly at the low-$k$ range, compared to averaged metrics, such as mean squared error.

## 5. Uncertainty of the ensemble model

Since five sub-models were trained independently using different random data partitions, the ensemble prediction for $c(k)$ is defined as:

$$c(k) = \frac{1}{M}\sum_{m=1}^{M} c^{(m)}(k) \qquad \text{Eq [S5.1]}$$

Here, $m \in [1, \ldots, M]$ indexes the sub-models, with $M = 5$. $c^{(m)}(k)$ is the predicted function from the $m$-th sub-model. The uncertainty at each $k$ is quantified by the standard deviation across sub-model predictions as follows:

$$\sigma_c(k) = \sqrt{\frac{1}{M}\sum_{m=1}^{M} [c^{(m)}(k) - c(k)]^2} \qquad \text{Eq [S5.2]}$$

The resulting $\sigma_c(k)$ values are shown in the gray area in **Figure 3A** in the main text to indicate the ensemble model uncertainty.

## 6. Investigation of model data efficiency

To evaluate the data efficiency of the ensemble model, we trained additional sub-models using various randomly selected fractions of the full dataset (e.g., 0.1, 0.2, 0.4, 0.6, 0.8, and 1.0). At each data fraction, we computed their individual mean sum of absolute residuals (MSAR) values for $g(r)$ across all training and validation state points. The MSAR values were then averaged over five sub-models to quantify an ensemble model's performance at each data fraction.

As shown in **Figure S8**, increasing the data fraction led to a near-polynomial decay in the average MSAR for both training and validation state points. The improvement plateaued near 80 % of the training data. We also evaluated the numerical stability of the ML closure as a function of dataset size and found that approximately 40 % of the full dataset (approximately 150 training-set state points) was required for the ensemble model to achieve a convergence rate of ~95 %.

## 7. Small-angle neutron scattering (SANS) experiments

Hydrogenated polystyrene (h-PS) was procured from Polymer Source[6]. The polymers were prepared by the vendor with an anionic polymerization process using a sec-butyl Li initiator and terminated by acidic methanol end groups. Hydrogenated (h) p-xylene (para-xylene) was purchased from Sigma Aldrich. Deuterated (d) p-xylene was purchased from Sigma Aldrich and Cambridge Isotopes. The polymers and solvents were used as received. All polymers were dissolved in p-xylene at 80 °C for a minimum of 1 hour prior to use.

Small-angle neutron scattering (SANS) experiments were performed at the NIST Center for Neutron Research in Gaithersburg, MD on the NGB 10 m and NG7 30 m SANS beamlines using standard configurations to encompass a broad range of wavenumbers $q$ ($0.003 < q < 0.6$) $Å^{-1}$. h-PS samples of 22,500 g/mol were dissolved in a 75:25 vol:vol mixture of deuterated and hydrogenated p-xylene, at polymer mass fractions 0.1 and 0.2. Using a polystyrene mass density of 1.05 $g/cm^3$, an (h) p-xylene density of 0.86 $g/cm^3$, and a (d) p-xylene density of 0.95 $g/cm^3$, the polymer mass fractions of 0.1 and 0.2 approximately correspond to polymer volume fraction, $\phi$, of 0.09 and 0.18, respectively. The polymer solutions were loaded into demountable cells with quartz windows (2 mm path length). The scattering length density of the h-PS was determined with contrast variation experiments to be $\rho_p = 1.99 \times 10^{-6} Å^{-2}$. The scattering intensity of the sample was placed on an absolute scale by measuring the empty beam flux and correcting for background, empty cell scattering, incoherent background and coherent contributions from the deuterated chain intramolecular contrast[7]. Data reduction was performed using the NIST Igor-based reduction algorithms[7].

The measured SANS intensity as a function of wavenumber, $I(q)$, was normalized by the corresponding volume fraction, $\phi$, in percent units as follows:

$$I(q) \rightarrow \frac{I(q)}{100 \cdot \phi} \qquad \text{Eq [S7.1]}$$

Each polymer chain in the experiments contained 220 monomers. To map this into our CG model, the CG chain length, $N$, was computed as follows:

$$N = \frac{220}{\sigma/\theta} \qquad \text{Eq [S7.2]}$$

where $\sigma = 11$ Å is the size of the CG bead in real units, chosen to match the approximate Kuhn segment length of the CG chains (1.5-1.8 times $\sigma$)[8] to that of polystyrene (PS, ~18 Å).[9] $\theta = 3$ Å is an approximate monomer diameter, as introduced in the main text.

The corresponding CG site number densities were calculated as follows:

$$\left.\begin{aligned} \phi &= \frac{N \cdot V_b}{V} \\ \rho &= \frac{N}{V} \end{aligned}\right\} \rightarrow \rho = \frac{\phi}{V_b} = \frac{\phi}{\frac{4}{3}\pi\left(\frac{\sigma}{2}\right)^3} \qquad \text{Eq [S7.3]}$$

where $V_b$ is the volume of a CG bead. $V$ is system volume. $\sigma = 1.0$. Using this relationship, $\phi = 0.09$ and $\phi = 0.18$ correspond to monomer number densities of $\rho = 0.17$ and $\rho = 0.34$, respectively.

## Supplemental Figures

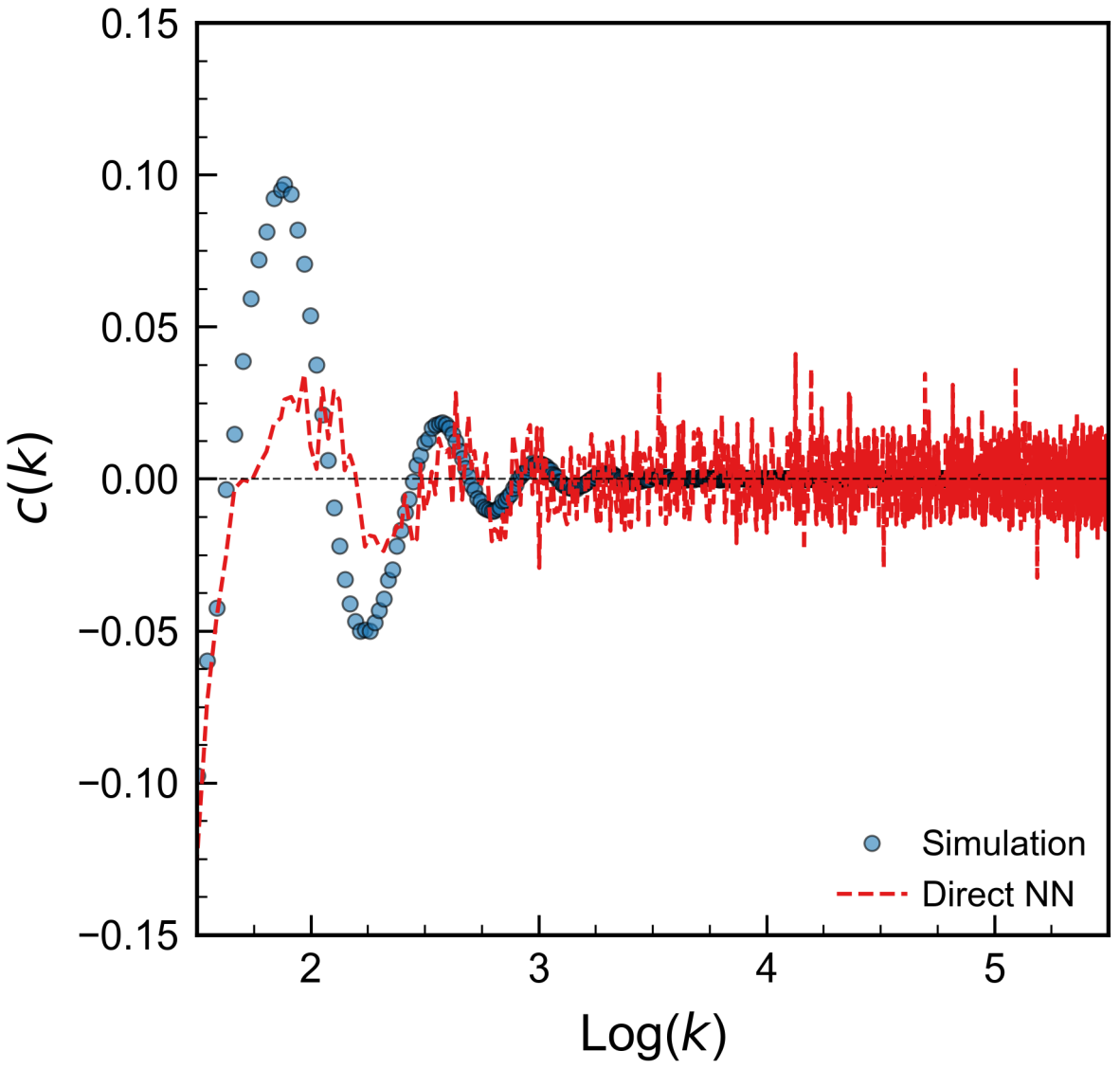

**SM Figure S1. Limitations of the direct neural network (NN) model.** A direct NN model prediction of $c(k)$ with no basis function expansion (red dashed curve) is compared to the simulation results (blue points) at the test state point of $N = 50$, $\varepsilon = 0.15$, $\rho = 0.40$.

**ALT TEXT:** Plot comparing the direct correlation function versus wavenumber for simulation data and a direct neural network model without a basis function dimensionality reduction.

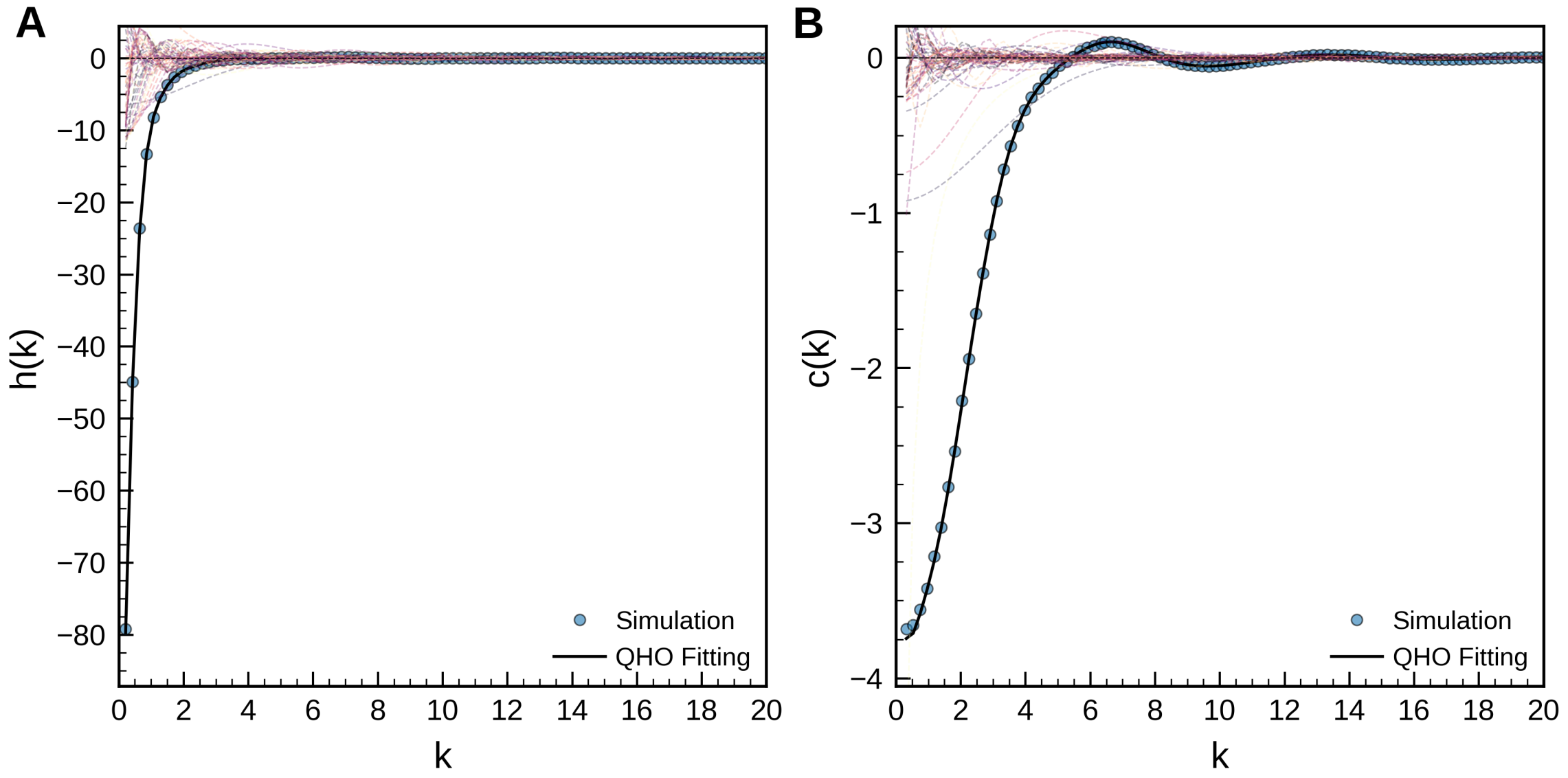

**SM Figure S2. Quantum harmonic oscillator (QHO) basis expansion of correlation functions.** (A) $h(k)$ and (B) $c(k)$ from the simulation (blue points) are decomposed into the QHO functions (colorful dashed curves) at the state point of $N = 50$, $\varepsilon = 0.10$, $\rho = 0.50$. The reconstructed functions (black solid curves) represent the linear sum of the QHO components.

**ALT TEXT:** Two-panel plot illustrating the quantum harmonic oscillator (QHO) basis function decomposition for the total correlation function in Panel A and direct correlation function in Panel B.

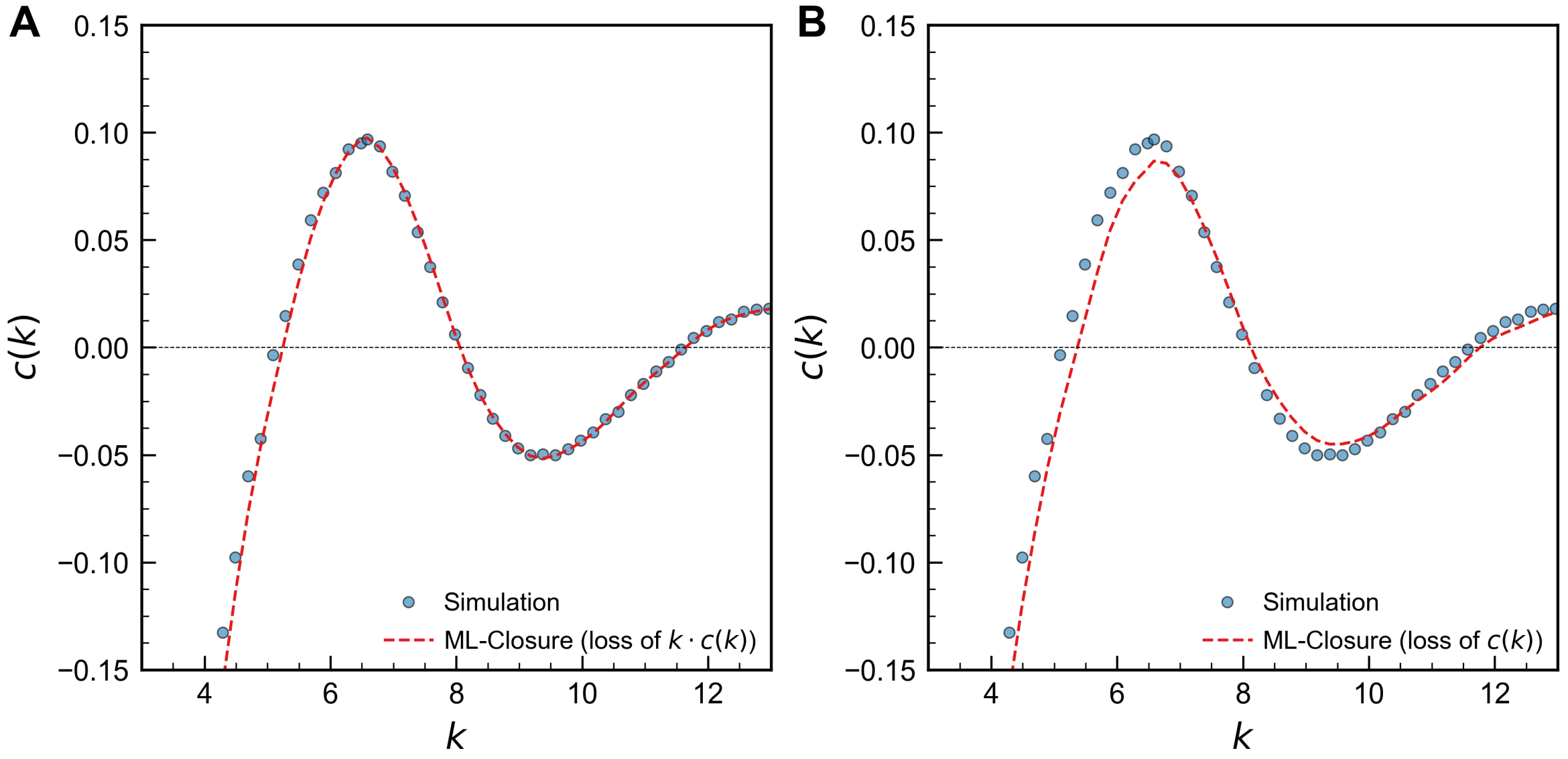

**SM Figure S3. Comparison of loss function targets during training.** In both panels, symbols denote simulation results, while red dashed curves represent the ML predictions at the state point of $N = 50$, $\varepsilon = 0.15$, $\rho = 0.40$. The model was trained using different loss function targets: (A) $k \cdot c(k)$ and (B) $c(k)$.

**ALT TEXT:** Two-panel plot comparing simulation results and ML closure predictions for the direct correlation function using different training loss function targets. Panel A uses $k \cdot c(k)$ as the loss target and Panel B uses $c(k)$ directly as the loss target.

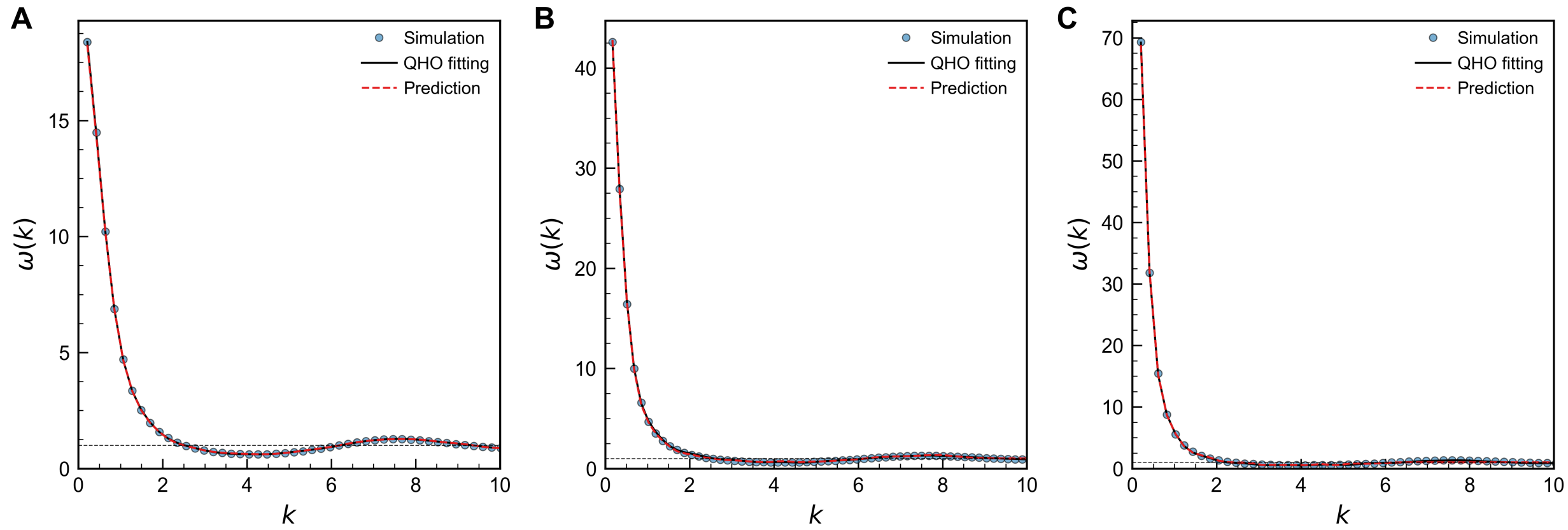

**SM Figure S4. Prediction examples from $\boldsymbol{\omega(k)}$ predictor.** Panels show representative cases at (A) $N = 20$, $\varepsilon = 0.30$, and $\rho = 0.25$; (B) $N = 50$, $\varepsilon = 0.15$, and $\rho = 0.25$; and (C) $N = 100$, $\varepsilon = 0.15$, and $\rho = 0.75$. Symbols represent simulation references, while red dashed curves show the corresponding predictions from the $\omega(k)$ predictor.

**ALT TEXT:** Three-panel plot displaying the performance of the ML model developed to predict the intramolecular correlation function across three representative state points.

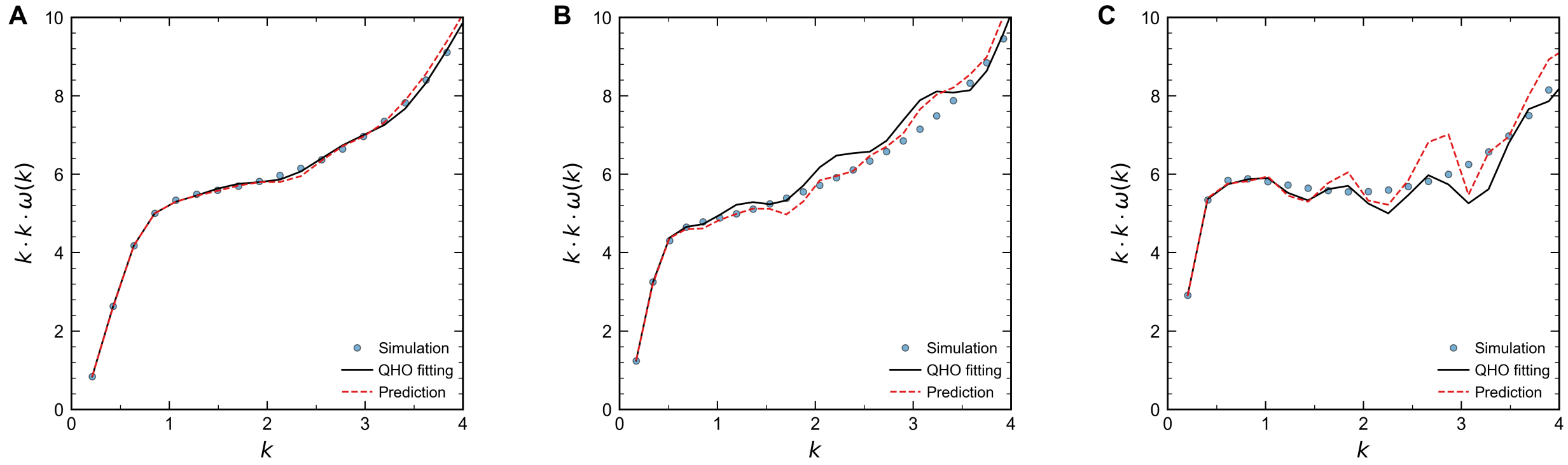

**SM Figure S5. Kratky representation for $\omega(k)$ predictor.** Same representative cases used in **SM Figure S4**. Symbols represent the simulation reference, while red dashed curves show the corresponding predictions from the $\omega(k)$ predictor.

**ALT TEXT:** Three panel figure showing the same data as SM Figure S4 but plotted in the Kratky plot representation, $k^2\omega(k)$ vs. $k$.

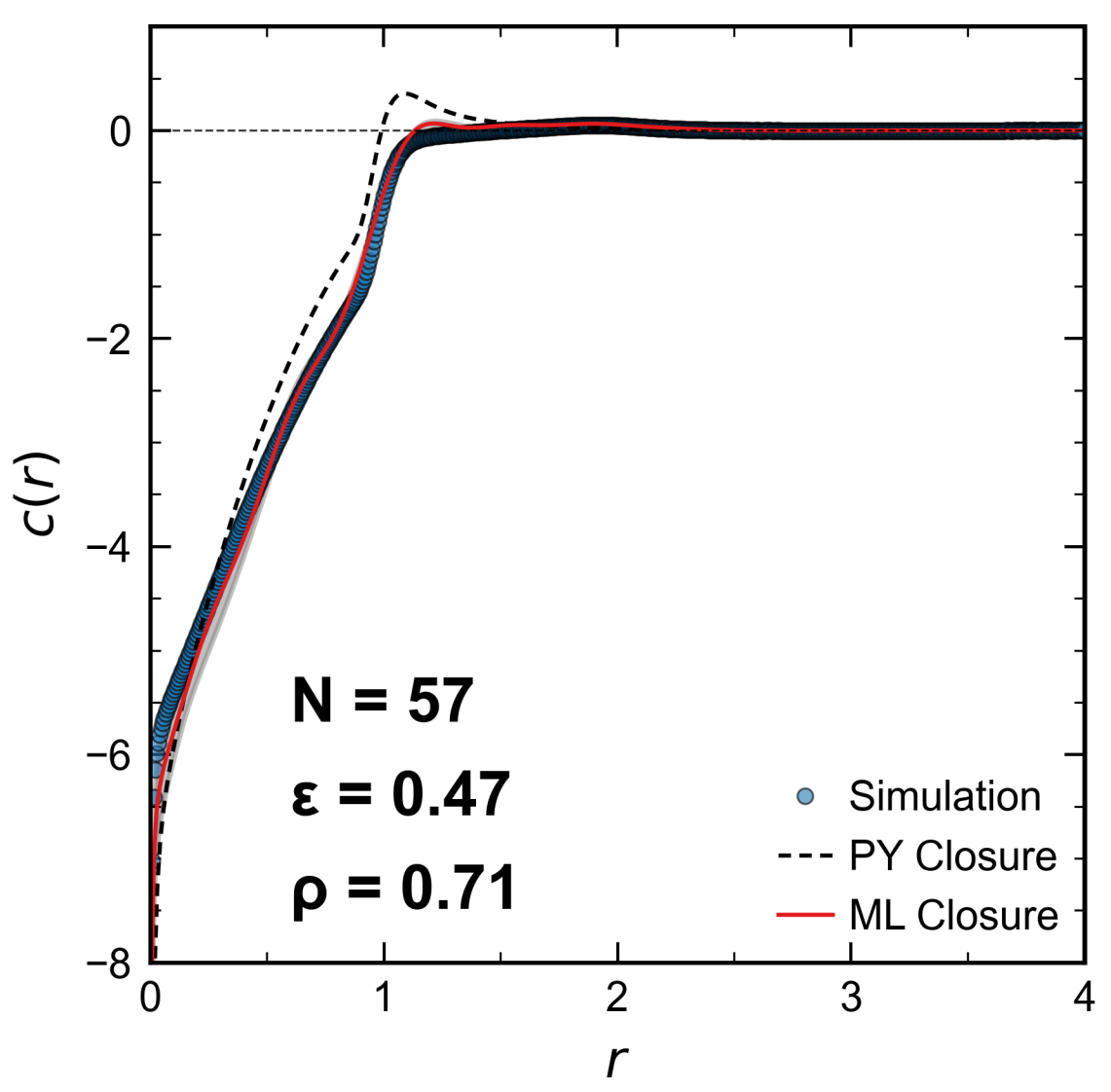

**SM Figure S6. Example direct correlation function in real space, $c(r)$.** Blue points: the simulation reference at the state point $N = 57$, $\varepsilon = 0.47$, $\rho = 0.71$. Red solid curve: the ML closure with the inter-model standard deviation (gray shaded area). Black dashed curve: the PY closure.

**ALT TEXT:** Plot comparing the real-space direct correlation function versus radial distance r for simulation data, the ML closure prediction, and the Percus-Yevick (PY) closure.

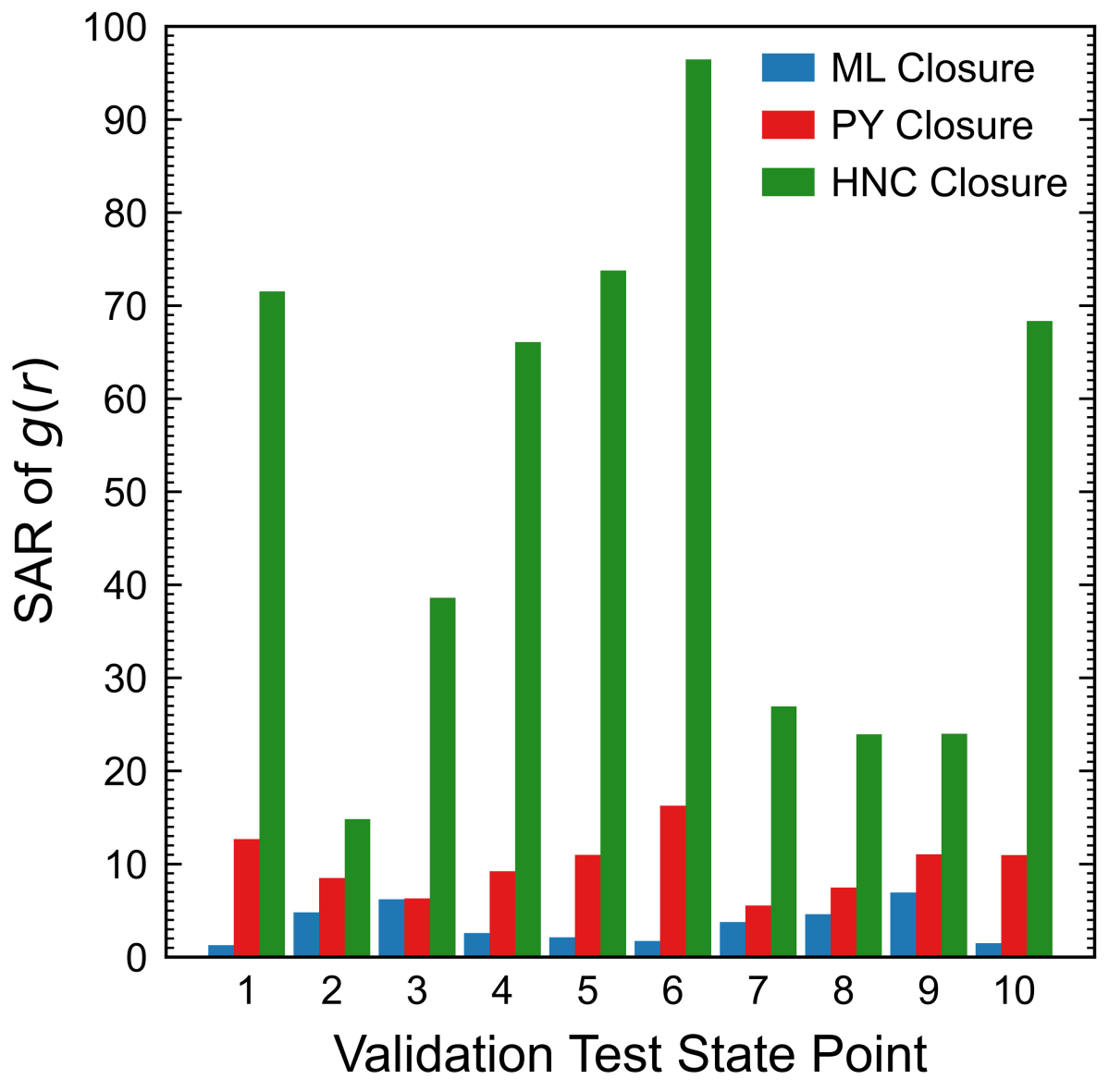

**SM Figure S7. Validation state points with HNC closure.** The sum of absolute residuals (SAR) of $g(r)$ at 10 validation state points for the ML closure (blue), PY closure (red), and HNC closure (green).

**ALT TEXT:** Bar chart of the sum of absolute residuals of the radial distribution function across 10 validation test state points for the ML closure, the PY closure, and the HNC closure.

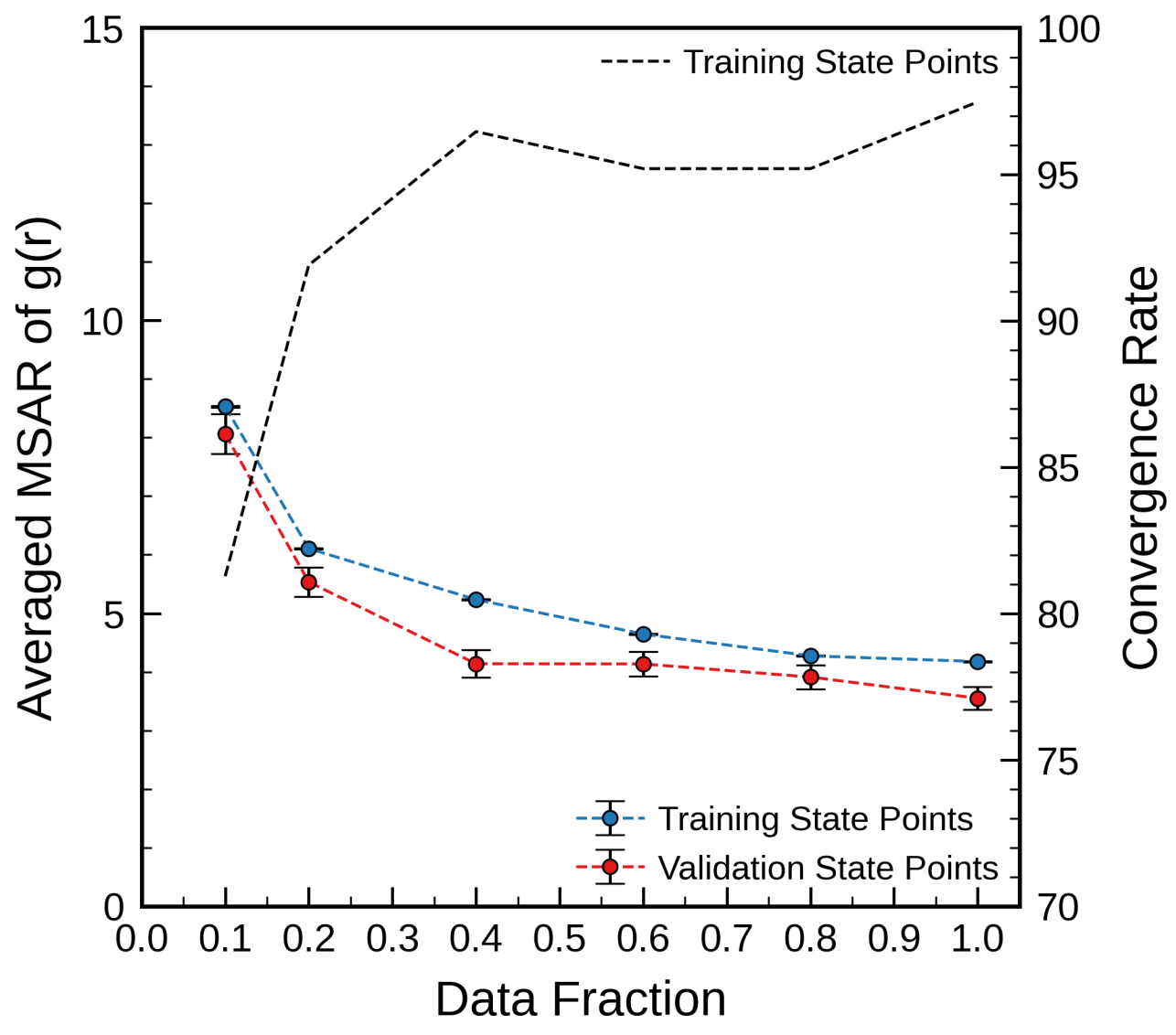

**SM Figure S8. Data efficiency of the ensemble model.** Blue and red symbols with dashed curves show the averaged MSAR for the training and validation state points, respectively, as a function of the dataset size. Error bars represent the standard error across the five-fold ensemble. The black dashed curve indicates the fraction of state points that achieved successful convergence at each data fraction.

**ALT TEXT:** Dual y-axis plot demonstrating model performance as a function of the amount of training data included in the ML closure model. The left axis shows the average mean sum of absolute residuals (MSAR) for the training and validation sets. The right axis shows the convergence rate of the PRISM iterative solution loop for differing dataset size.

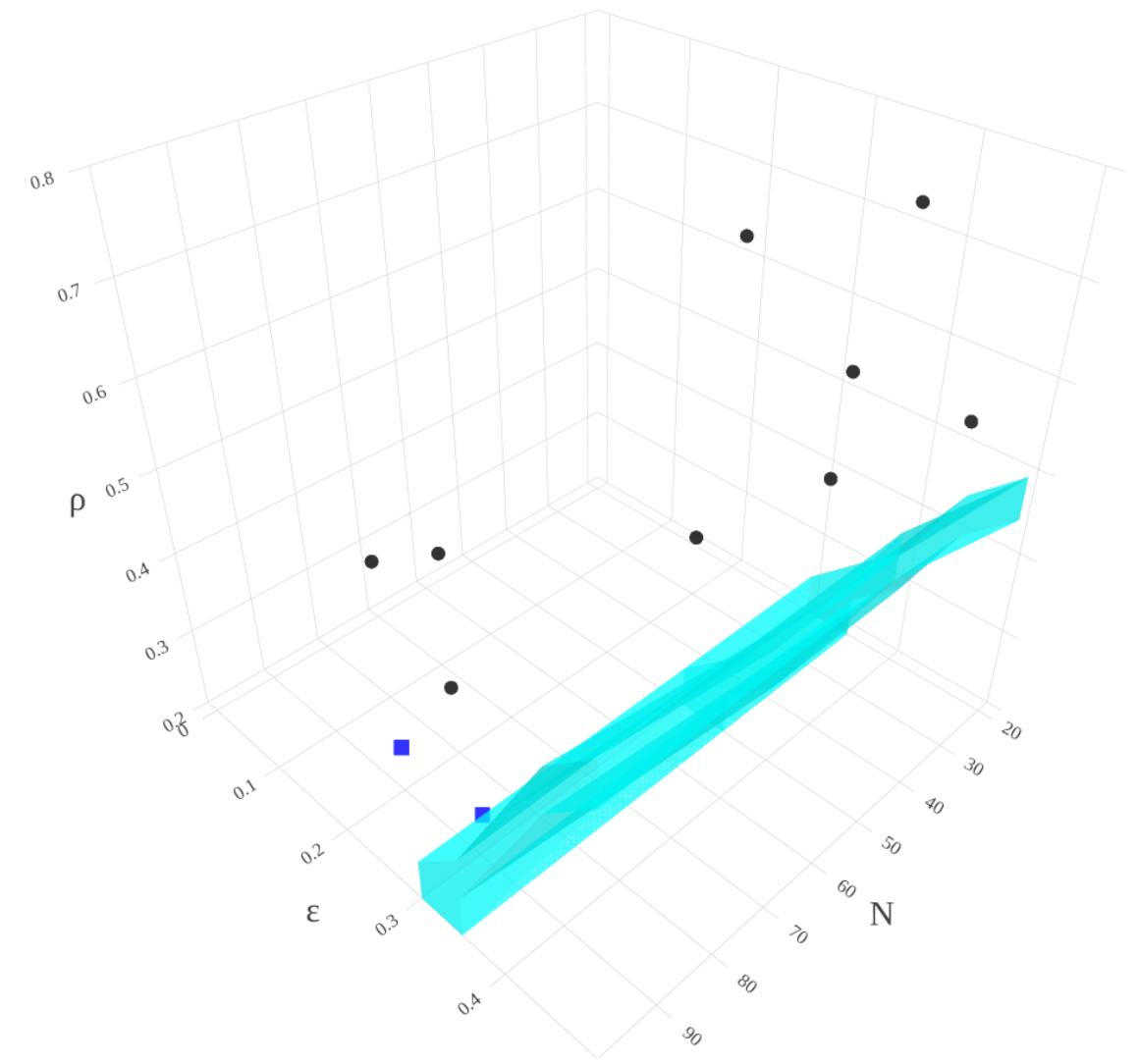

**SM Figure S9. Phase separation boundary and validation state points.** The light-blue region indicates the phase separation boundary determined from MD simulations. Points denote 11 randomly selected validation state points in the search space of $N$, $\varepsilon$, and $\rho$. Blue points highlight two specific cases: one on the edge of the phase separation boundary at $\varepsilon = 0.33$, and one well inside of the boundary at $\varepsilon = 0.23$.

**ALT TEXT:** Three-dimensional scatter plot in the $N$, $\varepsilon$, and $\rho$ parameter space showing the simulated phase separation boundary as a cyan shaded surface. Eleven validation state points are plotted as markers; two highlighted blue square markers demonstrate test cases located right on the edge ($\varepsilon = 0.33$) and well within ($\varepsilon = 0.23$) the single-phase region.

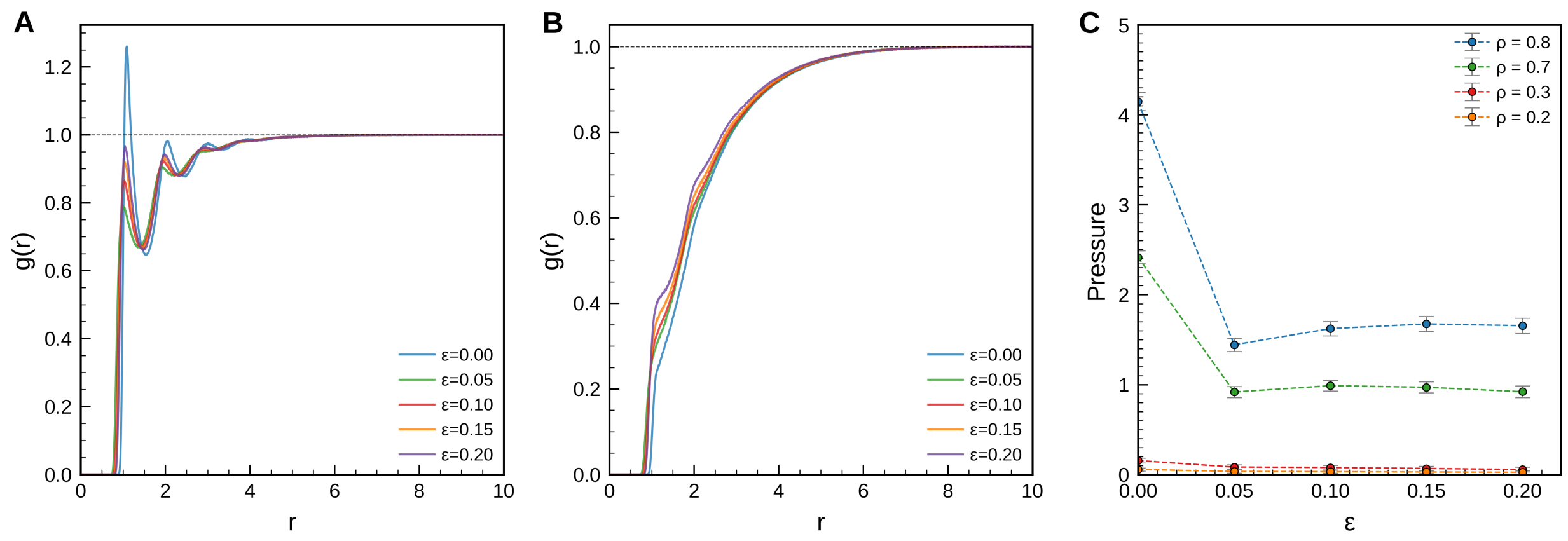

**SM Figure S10. Distinct behavior of the WCA potential.** Simulation results of $g(r)$ at various $\varepsilon$ values are shown for (A) $\rho = 0.80$, $N = 20$ and (B) $\rho = 0.20$, $N = 20$. Panel (C) shows the averaged pressure at $N = 20$ as a function of $\varepsilon$ and $\rho$, with error bars representing the standard deviation across simulation production data. In all plots, $\varepsilon = 0$ corresponds to the WCA potential.

**ALT TEXT:** Three-panel figure highlighting the unique behavior of the Weeks-Chandler-Andersen (WCA) potential. Panels A and B show the radial distribution function for high and low densities; at high density the WCA potential exhibits a prominent sharp first peak not present for weak attractive potentials. Panel C shows system pressure vs $\varepsilon$, revealing a sharp drop from WCA to LJ $\varepsilon$ = 0.05 before stabilizing at higher $\varepsilon$ values.

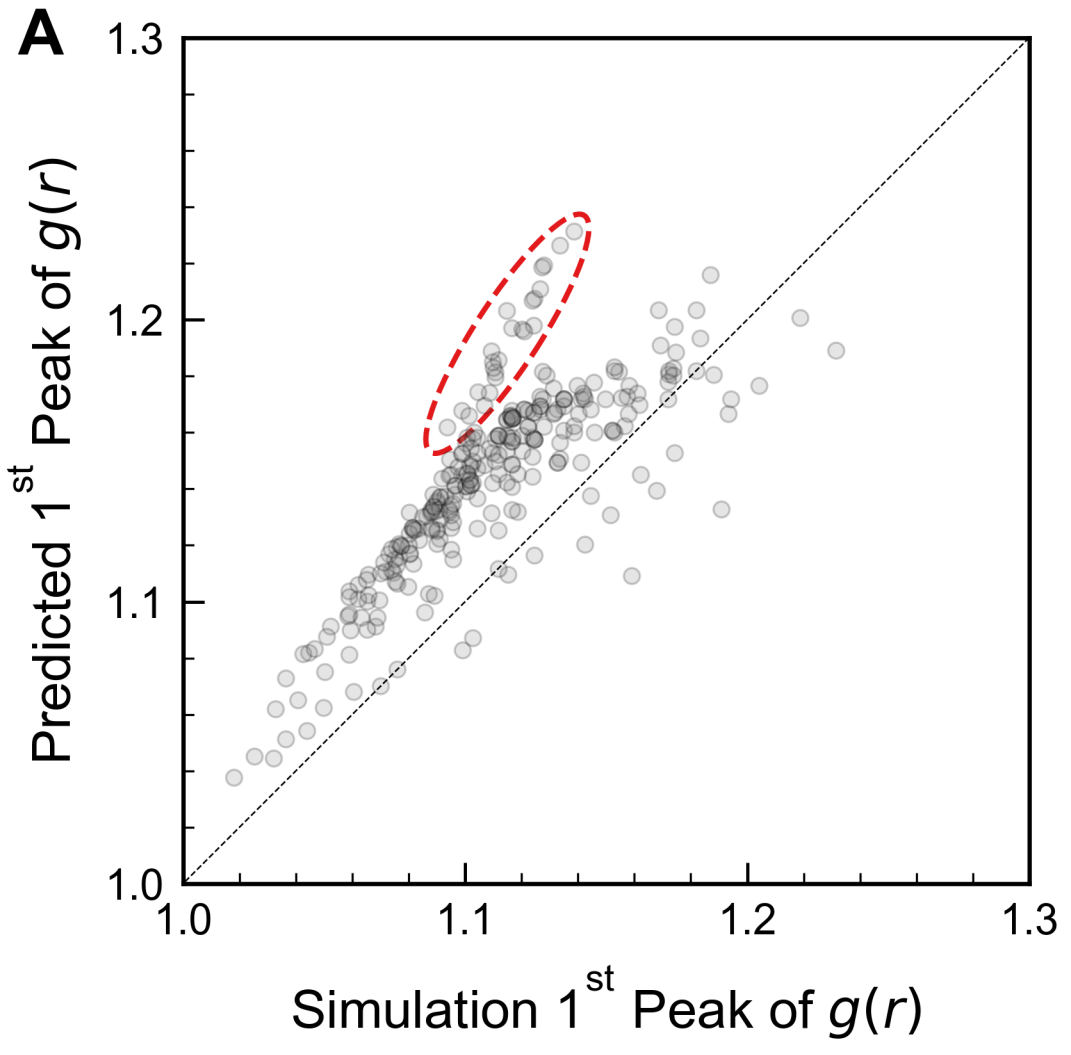

**SM Figure S11. Impact of the $LJ_{flag}$ feature on model performance.** Panels (A) and (B) compare the simulation results and ML closure predictions with and without the $LJ_{flag}$ feature, respectively. Red dashed circles indicate the state points that used the WCA potential ($\varepsilon = 0$).

**ALT TEXT:** Two scatter plots comparing the predicted first peak location of the radial distribution function with and without the $LJ_{flag}$ feature in the ML model.

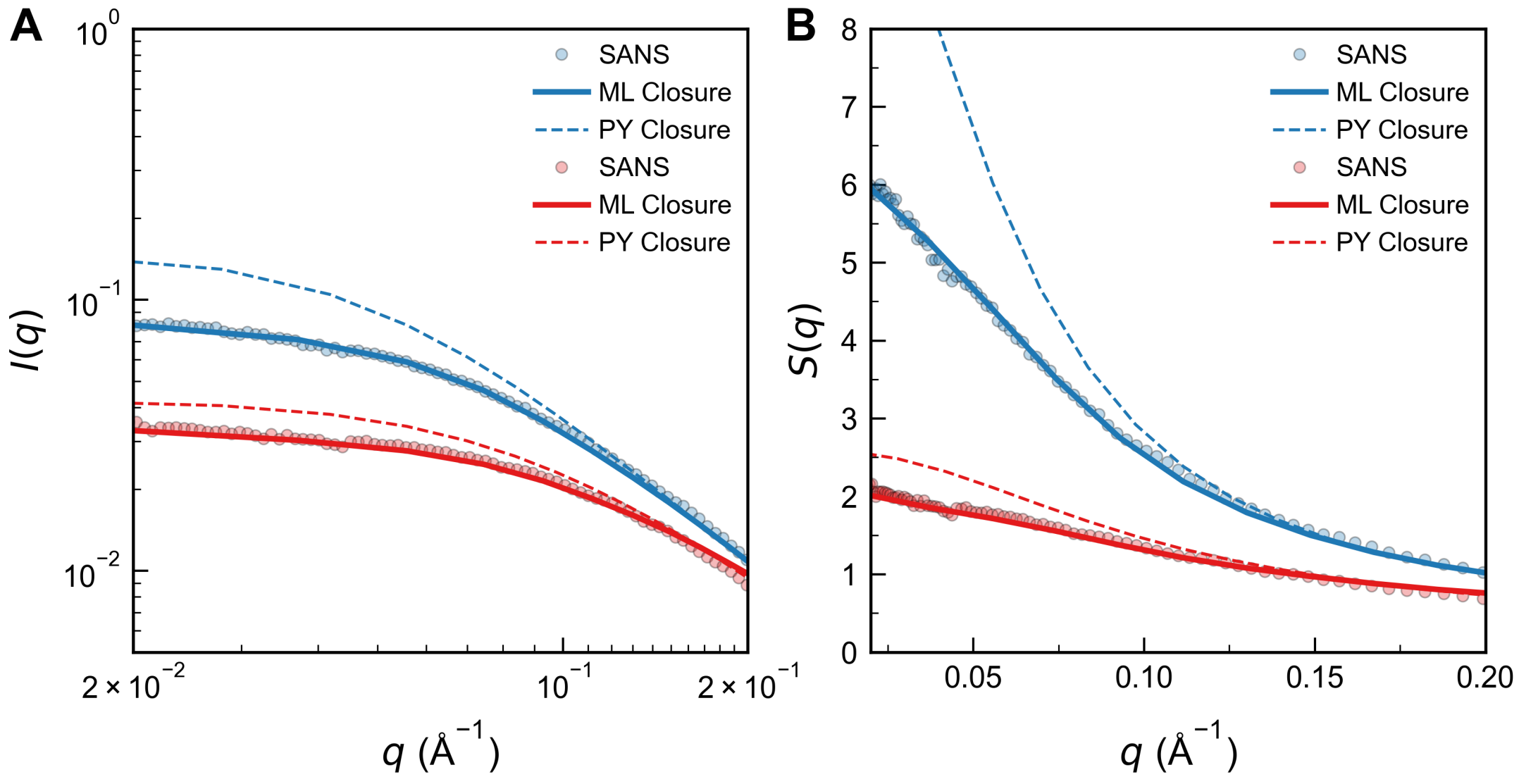

**SM Figure S12. ML closure results using simulated $\omega(k)$ at SANS conditions.** Panels (A) and (B) compare the fitted ML closure (solid curves) and PY closure (dashed curves) to the SANS experimental data (points) for $I(q)$ and $s(q)$, respectively, using the conditions given in Figure 5 of the main text.

**ALT TEXT:** Two-panel plot comparing ML closure and PY closure predictions to experimental SANS data for polystyrene solutions at two polymer volume fractions. Panel A presents scattering intensity vs wavenumber, and Panel B presents structure factor vs wavenumber.

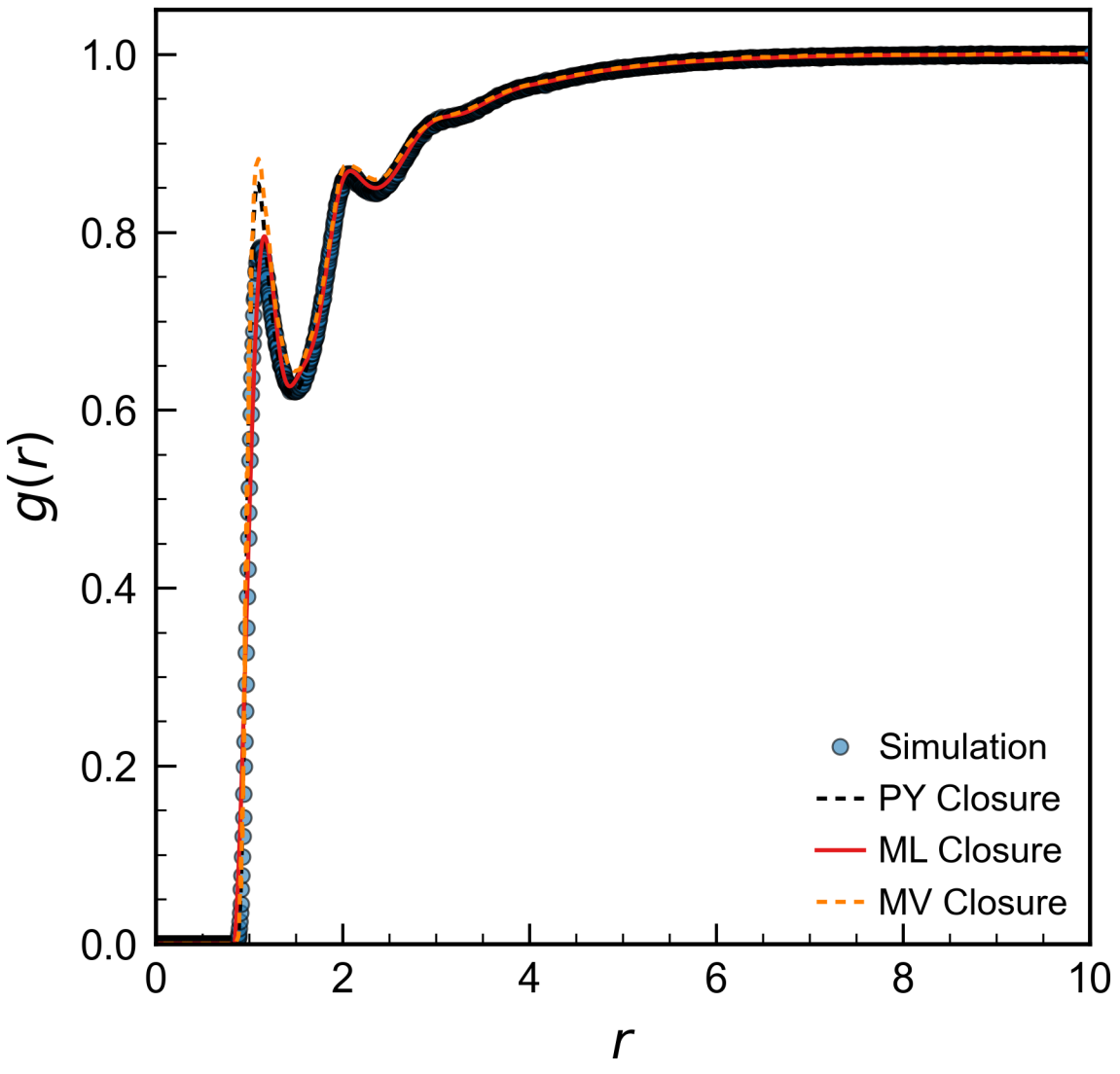

**SM Figure S13. Comparison of the Modified Verlet (MV), ML, and PY closures.** The pair correlation function, $g(r)$, is shown for a randomly chosen state point ($N = 26$, $\varepsilon = 0.46$, $\rho = 0.56$).

**ALT TEXT:** Radial distribution function plot comparing simulation data against the ML closure, the Percus-Yevick closure, and the Modified Verlet MV closure.